\documentclass[aps,prc,twocolumn,superscriptaddress,showpacs]{revtex4}
\usepackage{subfigure}
\usepackage{amsmath,epsfig}
\usepackage{graphicx}
\usepackage{dcolumn}
\usepackage{bm}
\usepackage{multirow}
\usepackage{hhline}
\begin{document}


\title{$K^{*0}$ production in Cu+Cu and Au+Au collisions at
      $\sqrt{s_{\mathrm{NN}}}$ = 62.4 GeV and 200 GeV}

\affiliation{Argonne National Laboratory, Argonne, Illinois 60439, USA}
\affiliation{University of Birmingham, Birmingham, United Kingdom}
\affiliation{Brookhaven National Laboratory, Upton, New York 11973, USA}
\affiliation{University of California, Berkeley, California 94720, USA}
\affiliation{University of California, Davis, California 95616, USA}
\affiliation{University of California, Los Angeles, California 90095, USA}
\affiliation{Universidade Estadual de Campinas, Sao Paulo, Brazil}
\affiliation{University of Illinois at Chicago, Chicago, Illinois 60607, USA}
\affiliation{Creighton University, Omaha, Nebraska 68178, USA}
\affiliation{Czech Technical University in Prague, FNSPE, Prague, 115 19, Czech Republic}
\affiliation{Nuclear Physics Institute AS CR, 250 68 \v{R}e\v{z}/Prague, Czech Republic}
\affiliation{University of Frankfurt, Frankfurt, Germany}
\affiliation{Institute of Physics, Bhubaneswar 751005, India}
\affiliation{Indian Institute of Technology, Mumbai, India}
\affiliation{Indiana University, Bloomington, Indiana 47408, USA}
\affiliation{Alikhanov Institute for Theoretical and Experimental Physics, Moscow, Russia}
\affiliation{University of Jammu, Jammu 180001, India}
\affiliation{Joint Institute for Nuclear Research, Dubna, 141 980, Russia}
\affiliation{Kent State University, Kent, Ohio 44242, USA}
\affiliation{University of Kentucky, Lexington, Kentucky, 40506-0055, USA}
\affiliation{Institute of Modern Physics, Lanzhou, China}
\affiliation{Lawrence Berkeley National Laboratory, Berkeley, California 94720, USA}
\affiliation{Massachusetts Institute of Technology, Cambridge, MA 02139-4307, USA}
\affiliation{Max-Planck-Institut f\"ur Physik, Munich, Germany}
\affiliation{Michigan State University, East Lansing, Michigan 48824, USA}
\affiliation{Moscow Engineering Physics Institute, Moscow Russia}
\affiliation{City College of New York, New York City, New York 10031, USA}
\affiliation{NIKHEF and Utrecht University, Amsterdam, The Netherlands}
\affiliation{Ohio State University, Columbus, Ohio 43210, USA}
\affiliation{Old Dominion University, Norfolk, VA, 23529, USA}
\affiliation{Panjab University, Chandigarh 160014, India}
\affiliation{Pennsylvania State University, University Park, Pennsylvania 16802, USA}
\affiliation{Institute of High Energy Physics, Protvino, Russia}
\affiliation{Purdue University, West Lafayette, Indiana 47907, USA}
\affiliation{Pusan National University, Pusan, Republic of Korea}
\affiliation{University of Rajasthan, Jaipur 302004, India}
\affiliation{Rice University, Houston, Texas 77251, USA}
\affiliation{Universidade de Sao Paulo, Sao Paulo, Brazil}
\affiliation{University of Science \& Technology of China, Hefei 230026, China}
\affiliation{Shandong University, Jinan, Shandong 250100, China}
\affiliation{Shanghai Institute of Applied Physics, Shanghai 201800, China}
\affiliation{SUBATECH, Nantes, France}
\affiliation{Texas A\&M University, College Station, Texas 77843, USA}
\affiliation{University of Texas, Austin, Texas 78712, USA}
\affiliation{Tsinghua University, Beijing 100084, China}
\affiliation{United States Naval Academy, Annapolis, MD 21402, USA}
\affiliation{Valparaiso University, Valparaiso, Indiana 46383, USA}
\affiliation{Variable Energy Cyclotron Centre, Kolkata 700064, India}
\affiliation{Warsaw University of Technology, Warsaw, Poland}
\affiliation{University of Washington, Seattle, Washington 98195, USA}
\affiliation{Wayne State University, Detroit, Michigan 48201, USA}
\affiliation{Institute of Particle Physics, CCNU (HZNU), Wuhan 430079, China}
\affiliation{Yale University, New Haven, Connecticut 06520, USA}
\affiliation{University of Zagreb, Zagreb, HR-10002, Croatia}

\author{M.~M.~Aggarwal}\affiliation{Panjab University, Chandigarh 160014, India}
\author{Z.~Ahammed}\affiliation{Lawrence Berkeley National Laboratory, Berkeley, California 94720, USA}
\author{A.~V.~Alakhverdyants}\affiliation{Joint Institute for Nuclear Research, Dubna, 141 980, Russia}
\author{I.~Alekseev~~}\affiliation{Alikhanov Institute for Theoretical and Experimental Physics, Moscow, Russia}
\author{J.~Alford}\affiliation{Kent State University, Kent, Ohio 44242, USA}
\author{B.~D.~Anderson}\affiliation{Kent State University, Kent, Ohio 44242, USA}
\author{DanielAnson}\affiliation{Ohio State University, Columbus, Ohio 43210, USA}
\author{D.~Arkhipkin}\affiliation{Brookhaven National Laboratory, Upton, New York 11973, USA}
\author{G.~S.~Averichev}\affiliation{Joint Institute for Nuclear Research, Dubna, 141 980, Russia}
\author{J.~Balewski}\affiliation{Massachusetts Institute of Technology, Cambridge, MA 02139-4307, USA}
% ...

\collaboration{STAR Collaboration}\noaffiliation


\begin{abstract}

We report on $K^{*0}$ production at mid-rapidity in Au+Au and Cu+Cu 
collisions at $\sqrt{s_{\mathrm{NN}}}$ = 62.4 and 200 GeV collected by the 
Solenoid Tracker at RHIC (STAR) detector. The $K^{*0}$ is reconstructed via 
the hadronic decays $K^{*0}\rightarrow K^ + \pi^-$ and $\overline{K^{*0}} 
\rightarrow K^- \pi^+$. Transverse momentum, $p_{\mathrm T}$, spectra are
 measured over a range 
of $p_{\mathrm T}$ extending from 0.2 GeV/$c$ up to 5 GeV/$c$. 
The center of mass energy and system size dependence of 
the rapidity density, $dN/dy$, and the average transverse momentum, 
$\langle p_{\mathrm T} \rangle$, are presented. The measured 
$N(K^{*0})/N(K)$ 
and $N(\phi)/N(K^{*0})$ 
ratios favor the dominance of re-scattering of decay 
daughters of $K^{*0}$ over the hadronic
regeneration for the $K^{*0}$ production. In the intermediate
$p_{\mathrm T}$ region ($2.0 < p_{\mathrm T} < 4.0$ GeV/$c$), the elliptic flow 
parameter, $v_{2}$, and the nuclear modification factor, $R_{CP}$, agree with 
the expectations from the quark coalescence model of particle production.
\end{abstract}
\pacs{ 25.75.Dw, 25.75.-q, 13.75.Cs}

\maketitle

\section{Introduction}
The main motivation for studying heavy-ion collisions at high energy 
is the study of Quantum Chromodynamics (QCD) in extreme conditions of 
high temperature and high energy density~\cite{whitepaper}. 
Ultra-relativistic nucleus-nucleus collisions at the Relativistic Heavy-Ion 
Collider (RHIC) create nuclear matter of high energy 
density over an extended volume, allowing QCD predictions to be tested
in the laboratory. At high temperature and density, QCD predicts a phase 
transition from nuclear matter to a state of deconfined
quarks and gluons known as the quark gluon plasma (QGP). One of the proposed 
signatures of the QGP state is the modification of 
vector meson production rates and their in-medium properties~\cite{medium1,medium2,strange}.

The $K^*$ meson is of particular interest due to its very short lifetime 
and its strange valence quark content. This makes the $K^*$ meson sensitive 
to the properties of the dense matter and strangeness production from 
an early partonic phase~\cite{Zhangbu,haibinPRC}. Since the 
lifetime of the $K^*$ is $\sim$ 4 fm/$c$, less than the lifetime of the system 
formed in heavy ion collisions~\cite{lifetime}, the $K^*$ is expected to decay, re-scatter, 
and 
regenerate all the way to the kinetic freeze-out (vanishing
elastic collisions). The characteristic properties of the resonance may be 
modified due to high density and/or high temperature of the medium 
causing in-medium 
effects. Various in-medium effects are partonic interaction 
with the surrounding matter, the interference between different 
scattering channels, and effects of phase space distortion due to 
re-scattering of the decay daughter particles \cite{rho PRL,phi_prl}.
Measurement of the $K^*$ meson properties such as mass, width, and yields at various
transverse momenta can provide insight for understanding the dynamics 
of the medium created in heavy-ion collisions.

Of particular interest in resonance production is understanding 
the role of re-scattering and regeneration effects. Due to the short 
$K^*$ lifetime, the pions and kaons from the $K^*$ that decay 
at the chemical freeze-out re-scatter with other hadrons. This would then 
inhibit the reconstruction of the parent $K^*$. 
However, in the presence of a large population 
of pions and kaons, these may scatter into a $K^*$ resonance state and 
thus contribute to the final measured yield~\cite{regeneration}. 
The interplay of 
these two competing processes 
becomes relevant for determining the $K^*$ yield in the hadronic medium. 
These processes depend on the time interval between chemical (vanishing 
inelastic collisions) and kinetic freeze-out, the source size, and 
the interaction cross section of the daughter hadrons. Since the $\pi \pi$ 
interaction cross section~\cite{pipicross} is larger than the 
$\pi K$ interaction cross section~\cite{piKcross}, 
the final observable $K^*$ yield may decrease compared to the primordial yield. 
A suppression of the yield ratio such as $N(K^*)/N(K)$ or $N(K^*)/N(\phi)$ 
is expected in 
heavy ion collisions compared to the same in $p + p$ collisions at similar
collision energies. This suppression can be used to set a lower limit on the time 
difference between the chemical and the kinetic freeze-out~\cite{rafelski,haibinPRC}. 
The experimental data on the system size, beam energy, and centrality 
dependence of this suppression can be used to correlate the lifetime of 
the fireball with its size. 
Although the measured values of the resonance yield, mean $p_{\mathrm T}$ and 
the elliptic anisotropy coefficient $v_{2}$ are all expected to be 
affected by collisional dissociation processes, semi-hard scattering 
and jet fragmentation,  the measurements presented in this paper have 
been discussed within the framework of re-scattering and regeneration.

The nuclear modification factors such as $R_{AA}$ and $R_{CP}$~\cite{rcp2} 
are of vital importance in differentiating between the effect of hadron 
mass and hadron type (baryon or meson) in the particle production. 
In the intermediate $p_{\mathrm T}$ range 
($2.0 < p_{\mathrm T} < 4.0$ GeV/$c$), $R_{CP}$ of $\Lambda$ (a baryon) and 
$K^{0}_S$ (a meson), as measured by STAR, are different. The observed differences 
are understood
as coming from differences in particle type (the baryon-meson effect), 
in agreement with the quark coalescence model \cite{rcp2,starpiprl}. Because 
the mass of the $K^*$ meson is comparable to the mass of 
the $\Lambda$ baryon, it is interesting to compare the $R_{CP}$ of $K^*$ with 
those of $K^{0}_S$ and $\Lambda$ to check whether the results confirm to the 
expectations of the quark coalescence model. Previous measurements of 
$R_{\mathrm {CP}}$ for $K^*$ in 
Au+Au collisions at $\sqrt{s_{\mathrm NN}}$ = 200 GeV were not precise 
enough to make such a conclusion~\cite{haibinPRC}. In this paper we present a 
measurement of $R_{CP}$ of the $K^*$ from a higher 
statistics data set collected in the year 2004.

In the intermediate $p_{T}$ range, the elliptic flow parameter, $v_2$, for 
different hadrons shows a deviation from the particle mass ordering as seen 
in the low $p_{T}$ regime ($p_{\mathrm T} < 1.5$ GeV/$c$)~\cite{vmass1,vmass2,rcp2}. 
For identified hadrons, $v_{2}$ follows a scaling with the number of 
constituent quarks, $n$, as expected from the quark coalescence model~\cite{phi_prl,rcp2}. 
The $K^{*}$ meson  is expected to follow the scaling law 
with $n=2$. The $K^{*}$ produced via regeneration of kaons and pions during 
hadronization, on the other hand, would follow the $n = 4$ scaling~\cite{nonaka}. 
Previous STAR measurements with a smaller data sample found $n=3 \pm 2$ 
\cite{haibinPRC} and could not conclusively determine the $K^{*}$ production 
mechanism. The additional $v_{2}$ data presented in this paper, for Au+Au collisions 
at 200 GeV, may conclusively 
provide information about the $K^{*}$ production mechanism in the 
intermediate $p_{T}$ range.

In previous STAR measurements, $K^{*}$ production was studied using data from
Au+Au, $p$+$p$, and $d$+Au collisions at 200 
GeV~\cite{haibinPRC,dAu} and Au+Au collisions at 130 GeV~\cite{Zhangbu}. The 
hadronic decay channels used in these analyses were  $K^{*0}\rightarrow K^ + \pi^-$, 
$\overline{K^{*0}} \rightarrow K^- \pi^+$, and $K^{*\pm}\rightarrow K^{0}_{S} + 
\pi^{\pm}$. In this paper we present new data on the $p_{T}$ distribution, 
$\langle p_{\mathrm T} \rangle$, and $dN/dy$ of $K^{*0}$ in Au+Au collisions at  
$\sqrt{s_{NN}}$ = 62.4 GeV and Cu+Cu collisions at $\sqrt{s_{NN}}$ = 62.4 GeV 
and 200 GeV. The data sample for Au+Au collisions at 200 GeV  
is 6.5 times larger than previous measurements, allowing us to make more 
quantitative conclusions from the $v_{2}$ and $R_{\mathrm {CP}}$ measurements. 
This broad systematic study, with two different colliding beam energies and 
two different colliding  species, enables us to  study the system 
size and energy dependence of various $K^{*0}$ properties in heavy ion 
collisions. To reduce statistical errors, the samples of $K^{*0}$ 
and $\overline{K^{*0}}$ were combined and are referred to as $K^{*0}$ 
in the present work, unless specified otherwise.

The paper is organized as follows. In Section II we discuss the detectors used 
in this analysis and details of the analysis procedure. 
For more details on the mixed event procedure used to extract the $K^{*0}$ yields, 
the systematic uncertainty estimation, and the procedure to obtain $v_{2}$, we 
refer the reader to our earlier publications~\cite{Zhangbu,haibinPRC}.
Our results on $p_{\mathrm T}$ 
spectra, $dN/dy$, 
$\langle p_{\mathrm T} \rangle$, particle ratios, $v_{2}$, and $R_{\mathrm {CP}}$
of $K^{*0}$ are presented in Section III. The results are summarized in Section IV.

\section{Experiment and Data Analysis}

The results reported here, represent data taken from Au+Au collisions
at $\sqrt{s_{NN}}$ = 62.4 GeV and 200 GeV in the year 2004 and Cu+Cu 
collisions at $\sqrt{s_{NN}}$ = 62.4 GeV and 200 GeV in the year 2005, 
using the STAR detector at RHIC~\cite{star}. The primary tracking
device within STAR, the Time Projection Chamber (TPC)~\cite{tpc}, was used 
for the track reconstruction of the decay daughters of $K^{*0}$. The TPC 
provides particle identification and momentum information of the charged 
particles by measuring their ionization energy loss, $dE/dx$~\cite{tpc}.

The data were collected with a minimum bias (MB) trigger. In Au+Au collisions 
the MB trigger requires a coincidence between two zero degree calorimeters (ZDC)~\cite{zdc}. 
The ZDCs are located 18 m away from the nominal collision point 
(center of TPC), in the beam direction, at  polar angle, $\theta$, less than $2$ mrad. 
For Cu+Cu collisions 
at 62.4 GeV, the minimum bias trigger was a combination of the signals from 
the ZDC and the Beam Beam Counter (BBC). The BBC at $3.3 < \eta < 5.0$ 
compensates for the trigger inefficiency of the ZDC in central events. 
To ensure uniform acceptance in the pseudo-rapidity, $\eta$, range studied, 
events with primary vertex position, $V_{Z}$, within $\pm$ 30 cm from the 
center of the TPC along the beam line were selected.

Centrality is defined as function of the fractional cross section measured 
as  function of the uncorrected charged particle multiplicity within the 
pseudo-rapidity window  $|\eta|<0.5$ for all events \cite{haibinPRC,starphiplb}. 
The most peripheral events were not taken into account due to large trigger 
and vertex finding inefficiencies. Table~\ref{Table:dataset} lists all the 
collision systems studied with the $V_{Z}$ cut, centrality range, and number of 
events used in the analysis. 
\begin{table}[htb]
\caption{List of datasets used in the analysis. Cuts on $V_{Z}$, centrality
range selected and number of events used are also given.} 
\label{Table:dataset}
\begin{tabular}{cccc}
\hline
\hline
  Collision systems  &  Centrality  & $|V_{Z}|$ cm  & Events \\
\hline
Au+Au~(62.4 GeV)   & 0-80\%  &  $<$ 30  & $~7\times 10^6$ \\

Cu+Cu~(62.4 GeV)   & 0-60\%  &  $<$ 30  & $~8\times10^6$ \\

Au+Au~(200 GeV)    & 0-80\%  &  $<$ 30   & $~13\times 10^6$  \\

Cu+Cu~(200 GeV)    & 0-60\%  &  $<$ 30  & $~19\times 10^6$ \\
\hline
\hline
\end{tabular}
\end{table}

\begin{figure}[htp]
\centering
\includegraphics[scale=0.4]{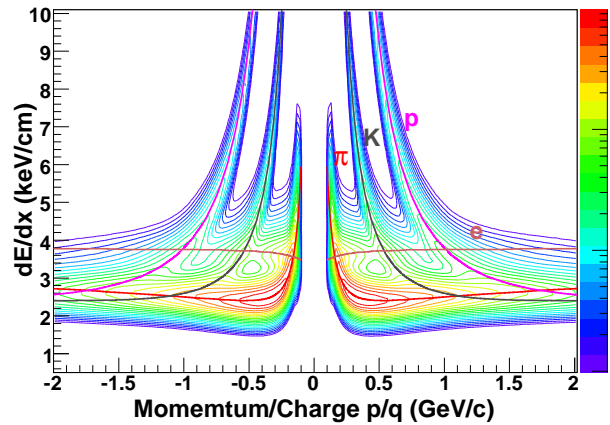}
\caption{(Color Online) $dE/dx$ for charged particles versus momentum 
divided by charge of the particle as measured in STAR TPC for Au+Au 
collisions at $\sqrt{s_{\mathrm {NN}}}$ = 200 GeV. The curves are the 
Bethe-Bloch predictions for different particle species.}
\label{dedx}
\end{figure}

Figure~\ref{dedx} shows the typical $dE/dx$ measured by the TPC in Au+Au collisions at 
200 GeV in year 2004 as a function of momentum, $p$, divided by charge of the particle, $q$. The different solid lines in the Figure~\ref{dedx} represent modified 
Bethe-Bloch predictions for different particle species~\cite{pdg,expecteddedx,spectralongpid}. 
The charged 
pions and kaons can be separated in momenta up to about 0.75 GeV/$c$ while (anti-)protons 
can be separated in momenta up to about 1.1 GeV/$c$. The particle identification 
can be quantitatively described by the variable $N_{\sigma}$, which for pions is 
defined as:
\begin{equation}
 N_{\sigma\pi}   =   \frac{1}{R}\log\frac{(dE/dx)_{measured}}{<dE/dx>_{\pi}}
\end{equation}
where $dE/dx_{measured}$ is the measured energy loss for a track, 
$ \langle dE/dx \rangle_{\pi}$ is the expected mean energy loss for a pion track at a 
given momentum~\cite{expecteddedx,spectralongpid}, and $R$ is the $dE/dx$ 
resolution which is around 8.1\%.
 
$K^{*0}$ mesons were reconstructed from their hadronic decay channels, $K^{*0}\rightarrow
K^ + \pi^-$ and $\overline{K^{*0}} \rightarrow K^- \pi^+$, using charged tracks
reconstructed with the TPC. 
Because the $K^{*0}$ decays within a very short time, its daughter particles seem 
to originate from the interaction point. Charged kaons and pions with a distance of 
closest approach to the primary vertex (DCA) less than 1.5 cm were considered
for Au+Au collisions at 62.4 GeV and Cu+Cu collisions at 62.4 and 200 GeV. In 
the case of 200 GeV Au+Au collisions, the DCA cut was set at 2.0 cm. 
The charged pion and kaon primary tracks thus selected were required to 
have their respective $N_{\sigma}$ values less than 2, with at least 15 fit points 
inside the TPC. This was done to ensure good track fitting with good momentum 
and $dE/dx$ resolution. Further, the ratio of the number of fit points to the number 
of maximum possible fit points was required to be greater than 0.55 to avoid 
selection of split tracks. In Au+Au collisions at 62.4 GeV and Cu+Cu collisions 
at 62.4 GeV and 200 GeV all the candidate tracks were 
required to have $|\eta| < 1$ while the tracks for the 200 GeV 
Au+Au collisions were 
required to have $|\eta| < 0.8$ to avoid the acceptance 
drop at high $\eta$ range. All tracks selected were also required to 
satisfy the condition that their $p_{\mathrm T}$ were greater than 
0.2 GeV/$c$. All the cuts used for the $K^{*0}$ analysis are summarized 
in Table~\ref{tab:Analysis Cuts}. 
\begin{table}[hbt]
\caption{List of track cuts for charged kaons and charged pions used in the 
$K^{*0}$ analysis in Au + Au and Cu+Cu collisions at 62.4 GeV and 200 GeV. 
NFitPnts is the number of fit points of a track in the TPC. MaxPnts is 
the number of maximum possible points of the track in the TPC.}
\label{tab:Analysis Cuts}
\begin{tabular}{ccc}
\hline
\hline
  Cut  & Values   \\
\hline
$N_{\sigma K}$           & (-2.0, +2.0) \\
$N_{\sigma \pi}$         & (-2.0, +2.0) \\
Kaon $p_{T}$ (GeV/$c$)     &(0.2, 10.0)\\
Pion $p_{T}$ (GeV/$c$)     &(0.2, 10.0) \\
NFitPnts                 & $>15$  \\
NFitPnts/Max Pnts        & $>0.55$ \\
Kaon and Pion $\eta$     & $|\eta| < 1.0$ \\
                         & $|\eta| < 0.8$ (Au+Au 200 GeV)\\
DCA                      & $< 1.5$ cm \\
                         & $< 2.0$ cm (Au+Au 200 GeV)\\
Pair Rapidity ($y$)      & $|y| <0.5$ \\
\hline
\end{tabular}
\end{table}
\begin{figure*}
\begin{center}
\includegraphics[scale=0.75]{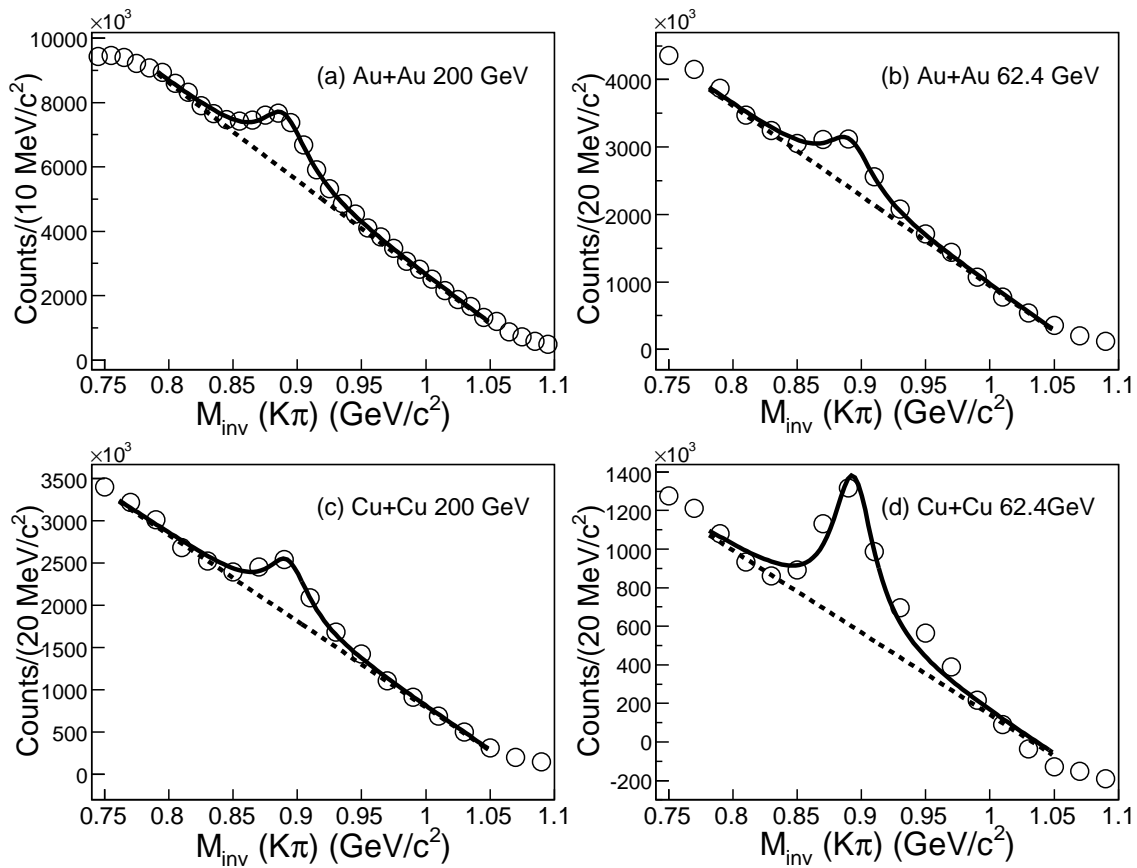}
\caption{The $K\pi$ pair invariant mass distribution integrated over the
$K^{*0}$ $p_{T}$ for minimum bias Au+Au and Cu+Cu collisions at 
$\sqrt{s_{\mathrm {NN}}}$ = 62.4 and  200 GeV after the mixed-event 
background subtraction. The solid curve is the signal fit to a Breit-Wigner 
function (Eqn. 2) + Linear function (Eqn. 3) 
while the dashed line is the linear function representing the residual background.}
\label{signal}
\end{center}
\end{figure*}

In a typical event, several hundred tracks originate from the 
primary collision vertex.  It is impossible to distinguish the tracks 
corresponding to the decay daughters of the $K^{*0}$ from other primary tracks. The 
$K^{*0}$ was reconstructed by calculating the invariant mass for each 
unlike-sign $K\pi$ pair in an event. The resultant distribution consists of 
the true $K^{*0}$ signal and contributions arising from random combination of 
unlike-sign  $K\pi$ pairs. The true $K^{*0}$ signal constitutes a very small 
fraction of the total invariant mass spectrum. The large random combinatorial 
background must be subtracted from the unlike-sign $K\pi$ 
invariant mass distribution to extract the $K^{*0}$ yield. This random 
combinatorial background 
distribution is obtained using the mixed-event 
technique \cite{mixevent1,mixevent2,dAu,starphiplb}. In the mixed event technique, the reference 
background distribution was built with uncorrelated unlike-sign $K\pi$ pairs 
from different events. For generating the mixed events, the data sample was 
divided into 10 bins in event multiplicity and 10 bins in $V_{Z}$. Unlike-sign 
$K\pi$ pairs from events having similar event multiplicity and $V_{Z}$ were 
selected for mixing. This was done to ensure that the characteristics 
of the mixed events generated were similar to the actual data. 
The generated mixed event 
sample was properly normalized to subtract the background from the same 
event unlike-sign invariant mass spectrum. The normalization factor 
was calculated by taking the ratio between the number of entries in the unlike-sign 
and the  mixed event distributions with invariant mass greater than 1.2 GeV/$c^2$. 
The $K\pi$ pairs are less likely to be correlated in this region.

\begin{figure}
\includegraphics[scale=0.4]{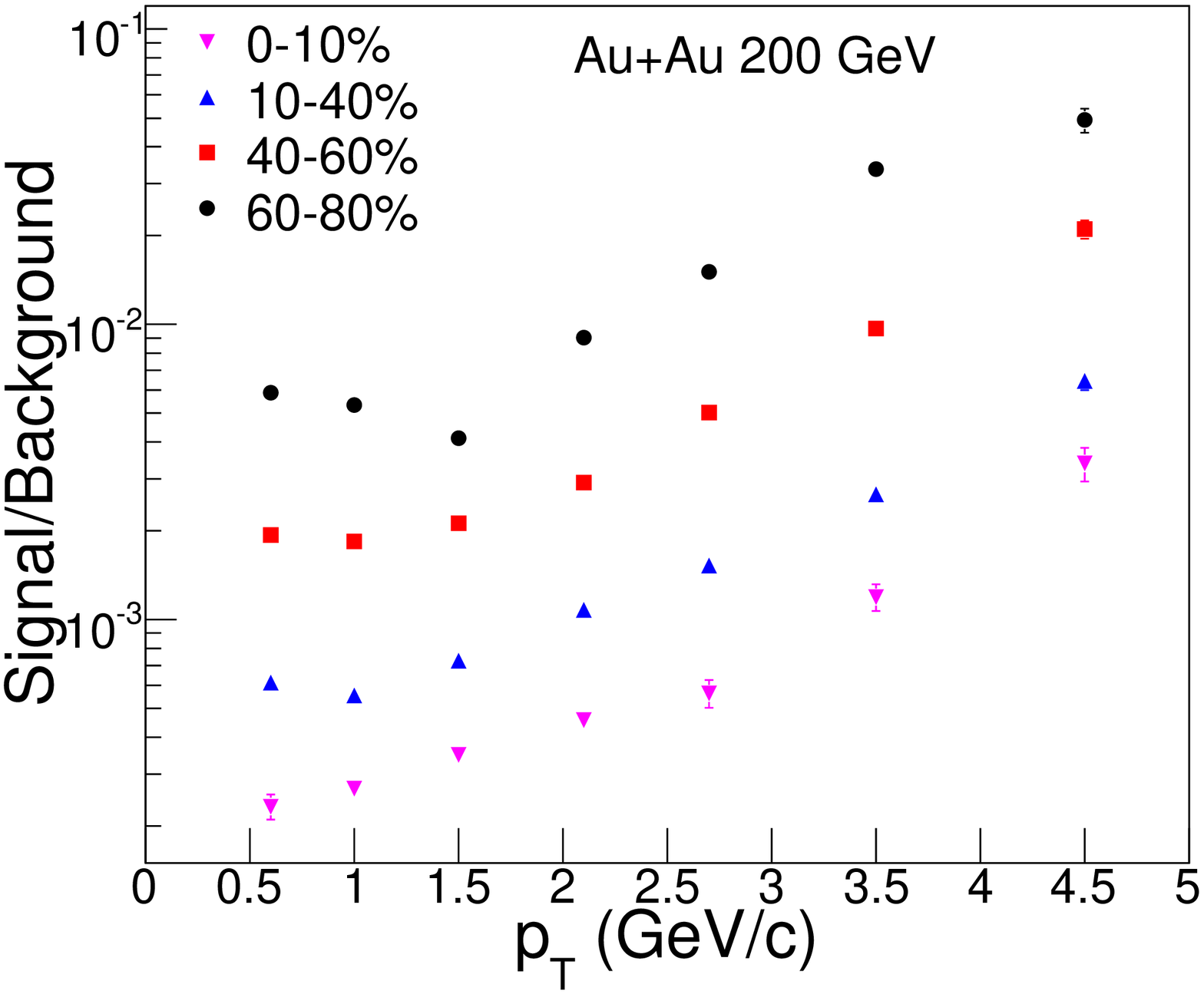}
\caption{The signal-to-background ratio for $K^{*0}$ measurements as a function
of $p_{\mathrm T}$ for different collision centrality bins in Au+Au collisions at 200 GeV.}
\label{sb}
\end{figure}

Figure~\ref{signal} shows the background-subtracted and $p_{\mathrm T}$-integrated 
unlike-sign $K\pi$ invariant mass spectra corresponding to minimum bias Au+Au and 
Cu+Cu collisions at $\sqrt{s_{\mathrm{NN}}}$ = 62.4 and 200 GeV. 
The signal-to-background ratio as a function of the $K\pi$  pair $p_{\mathrm T}$ for 
Au+Au collisions at $\sqrt{s_{\mathrm{NN}}}$ = 200 GeV is shown in Figure~\ref{sb}. 
The values are of similar order for other collision systems.
The signal-to-background ratio, S/B, is observed to increase with decreasing 
multiplicity of the events and shows
an increase with increasing $p_{\mathrm T}$. 
In the unlike-sign spectrum we also could have higher and/or lower $K\pi$ mass resonant 
states and non-resonant correlations due to particle misidentification and effects 
from elliptic flow in non-central collisions. These effects contribute 
significantly to the residual correlations near the signal~\cite{haibinPRC} 
that are not present in the mixed-event sample. These residual 
correlations are also subtracted from the background using a background 
function described in Section III .

\section{Results}

\subsection{$M_{K\pi}$ peak and width}

The invariant mass distributions (typical distributions shown in 
Figure~\ref{signal}) for various $p_{\mathrm T}$ bins, in Au+Au and 
Cu+Cu collisions at $\sqrt{s_{\mathrm{NN}}}$ = 62.4 and 200 GeV were fit 
using a function representing a non-relativistic Breit-Wigner (BW) 
shape plus a linear residual background (RBG).
The BW  and the RBG parts are as given below.
\begin{equation}
BW  = \frac{\Gamma_{0}}{(M_{K\pi} - M_{0})^{2} + \frac{\Gamma_{0}^{2}}{4}} 
\end{equation}
\begin{equation}
 RBG = a + bM_{K\pi}
\end{equation}
In the above equations, $M_{0}$ and $\Gamma_{0}$ are the mass and width of the 
$K^{*0}$; $a$ and $b$ are the intercept and slope for the linear residual 
background. 

\begin{figure}
\begin{center}
\includegraphics[scale=0.4]{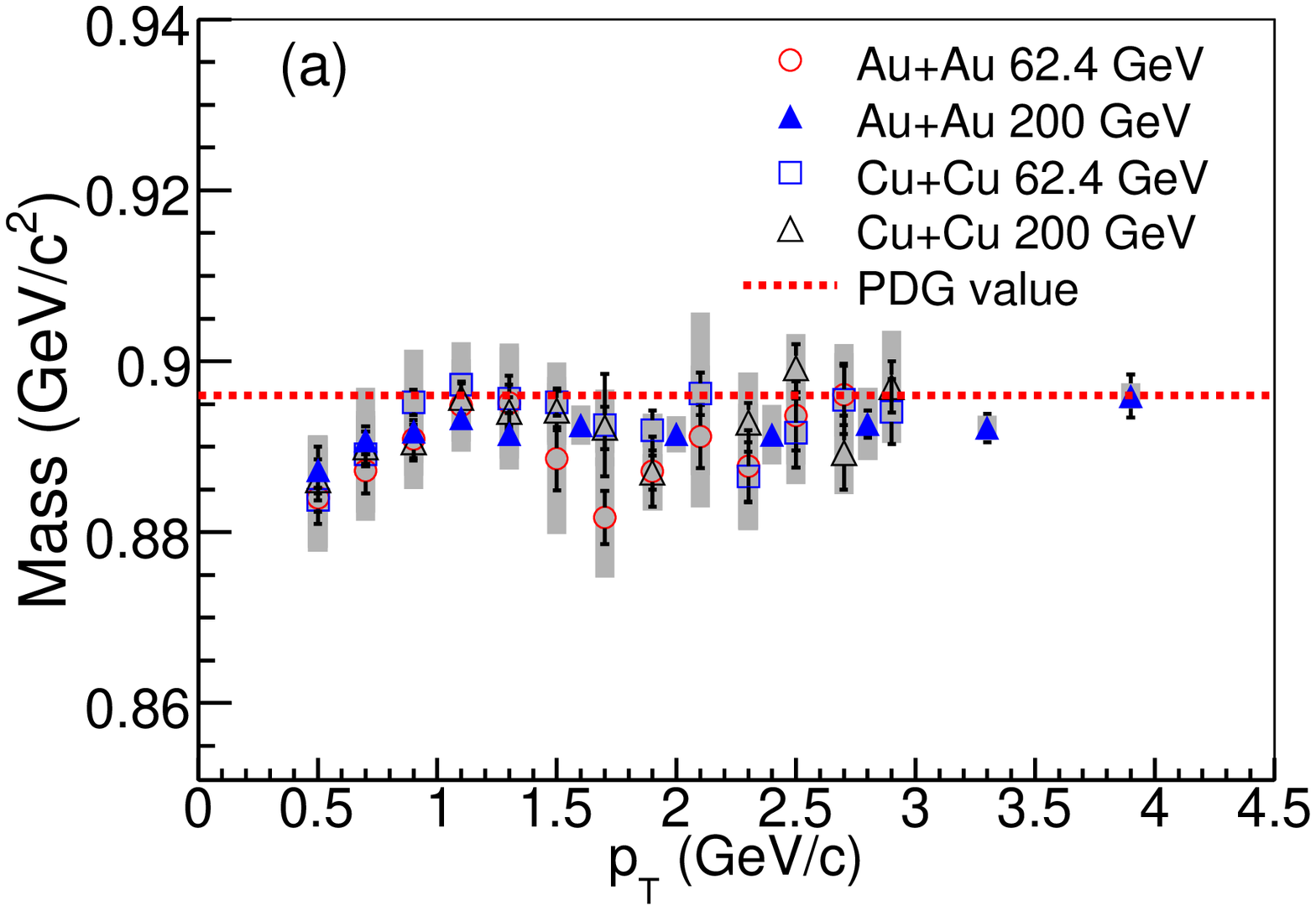}
\includegraphics[scale=0.4]{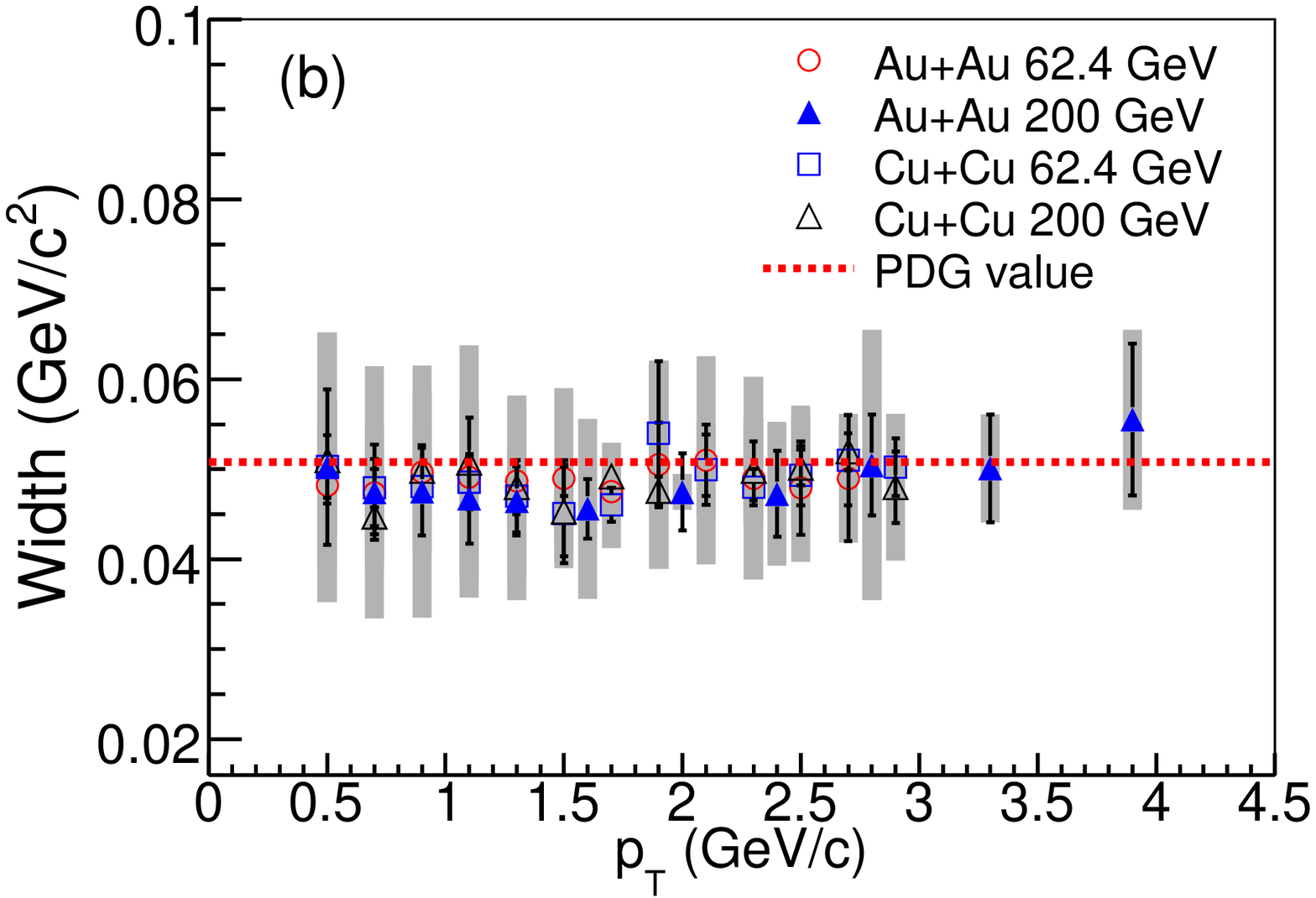}
\caption{$K^{*0}$ (a) mass and (b) width as a function of $p_{T}$ for minimum bias 
Au+Au and Cu+Cu collisions at $\sqrt{s_{NN}}$ = 62.4 and 200 GeV . The dashed 
line represents the PDG values of 896.0 MeV/$c^2$ and 50.3 MeV/$c^2$ for mass 
and width respectively.}
\label{masswidth}
\end{center}
\end{figure}
The variations of $M_{0}$ and $\Gamma_0$ with 
$p_{\mathrm T}$ are shown in Figure~\ref{masswidth}. The error bars shown correspond to 
statistical uncertainties while the bands represent systematic uncertainties. In the 
low $p_{\mathrm T}$ region ($<1$ GeV/$c$), the measured widths are consistent 
with the Particle Data Group (PDG) value of 50.3 MeV/$c^2$ 
while the measured masses are within 2$\sigma$ of the PDG value of 896.0 
MeV/$c^2$ \cite{pdg}. In the higher $p_{\mathrm T}$ range ($>$ 1 GeV/$c$), both the
mass and width of $K^{*0}$ are seen to be consistent with the PDG values.  
We do not observe any significant dependence of $K^{*0}$ mass and width on beam 
energy and colliding ion species studied. 
The systematic uncertainties on the $K^{*0}$ mass and width measurement were evaluated 
bin-by-bin, as a function of $p_{\mathrm T}$: (1) an uncertainty on the signal fit was 
evaluated by replacing the non-relativistic BW function with a relativistic BW function, 
(2) an uncertainty on the background was evaluated by varying residual background 
functions by using higher order polynomials, and (3) an uncertainty 
on track selection was calculated by varying the 
particle identification criteria and different cuts on the daughter tracks. In the 
above analysis, low $p_{\mathrm T}$ kaon tracks were corrected for energy loss due 
to multiple scattering in the detector \cite{starphiplb,spectralongpid}.

\begin{figure}
\begin{center}
\includegraphics[scale=0.4]{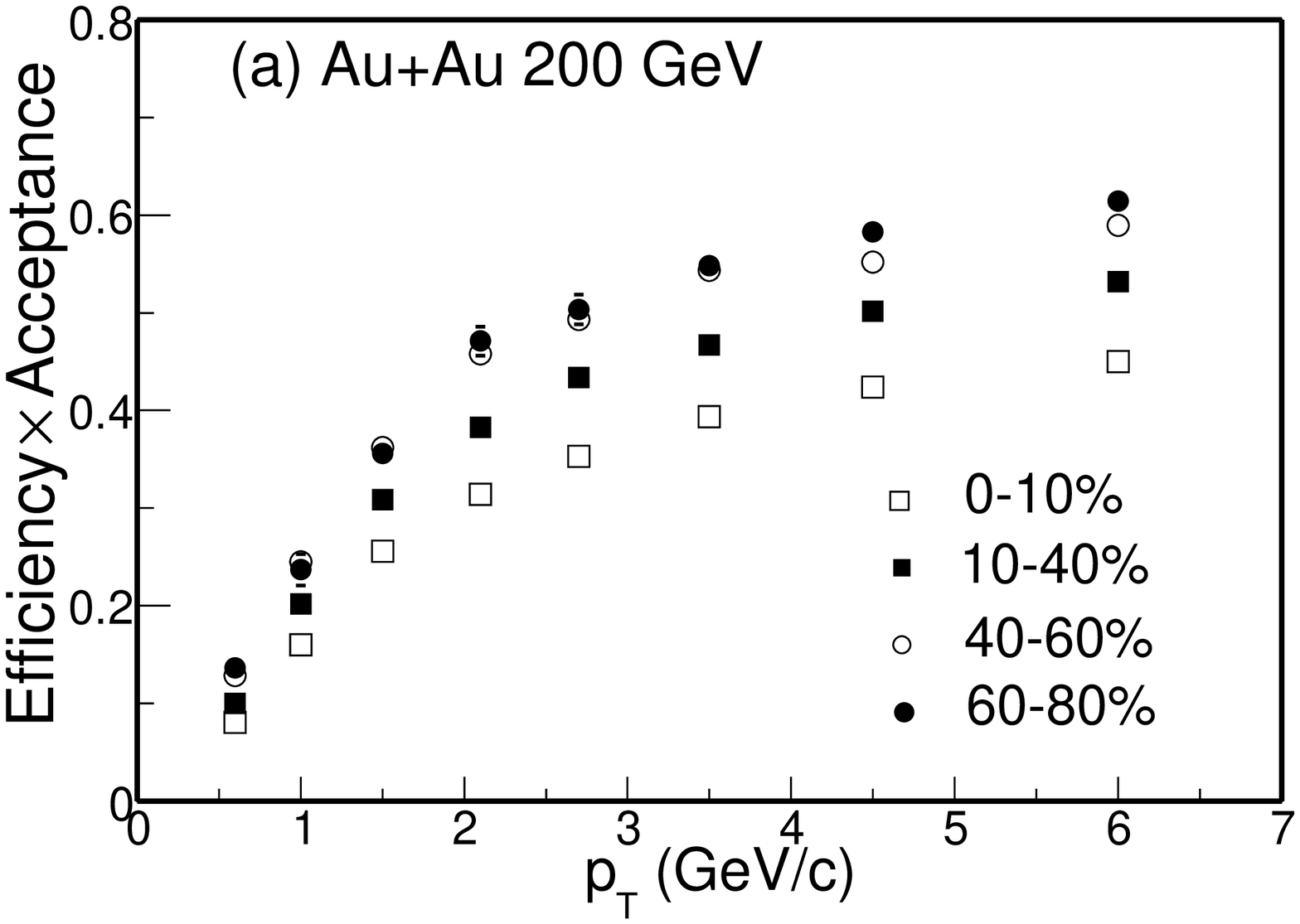}
\includegraphics[scale=0.45]{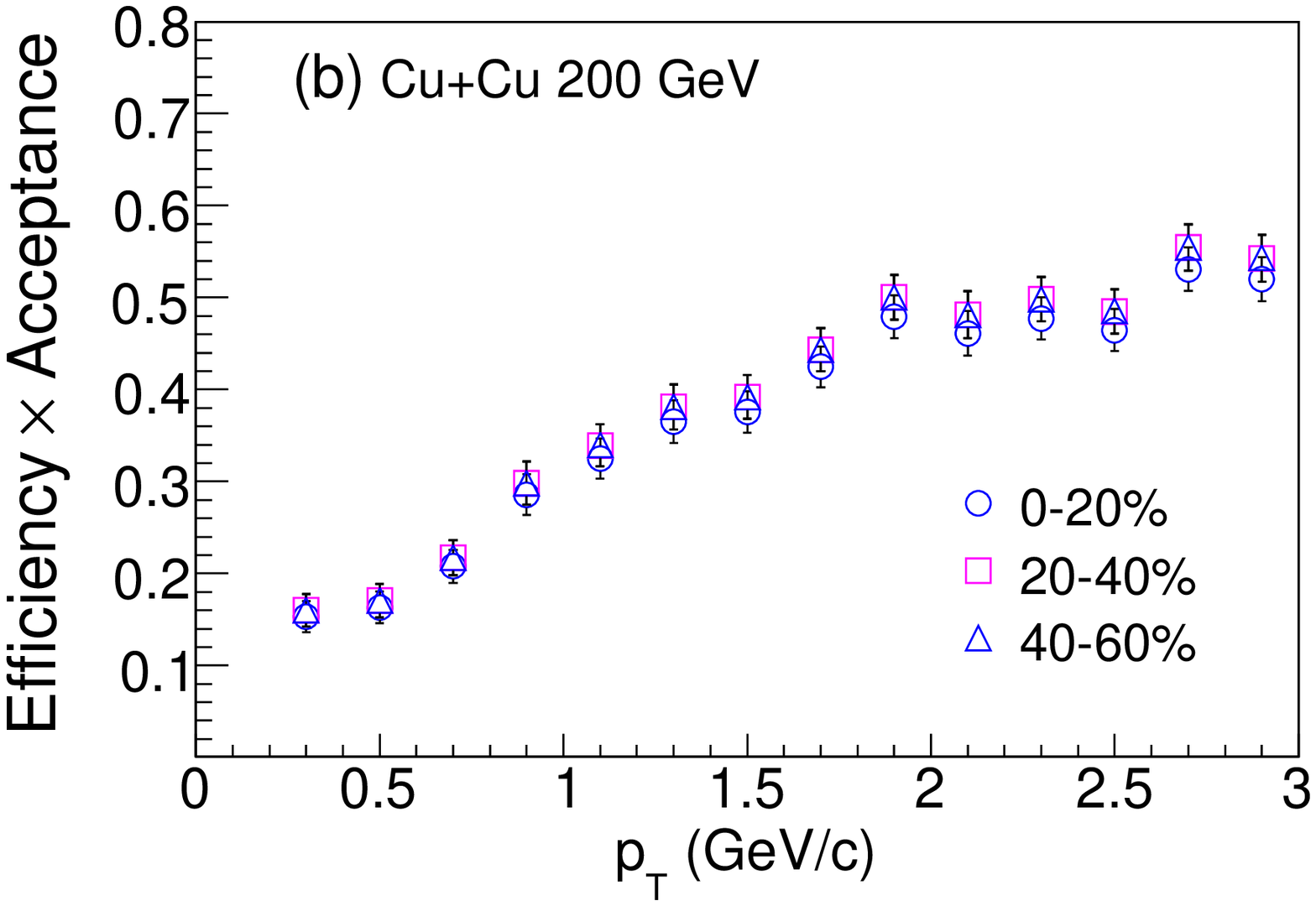}
\caption{The $K^{*0}$ reconstruction efficiency multiplied by the detector
acceptance as a function of $p_{\mathrm T}$ in (a) Au+Au ($|\eta| < 0.8$) and 
(b) Cu+Cu ($|\eta| < 1.0$) collisions at 200 GeV 
for different collision centrality bins.}
\label{eff}
\end{center}
\end{figure}

\begin{figure*}[htp]
\begin{center}
\includegraphics[scale=0.4]{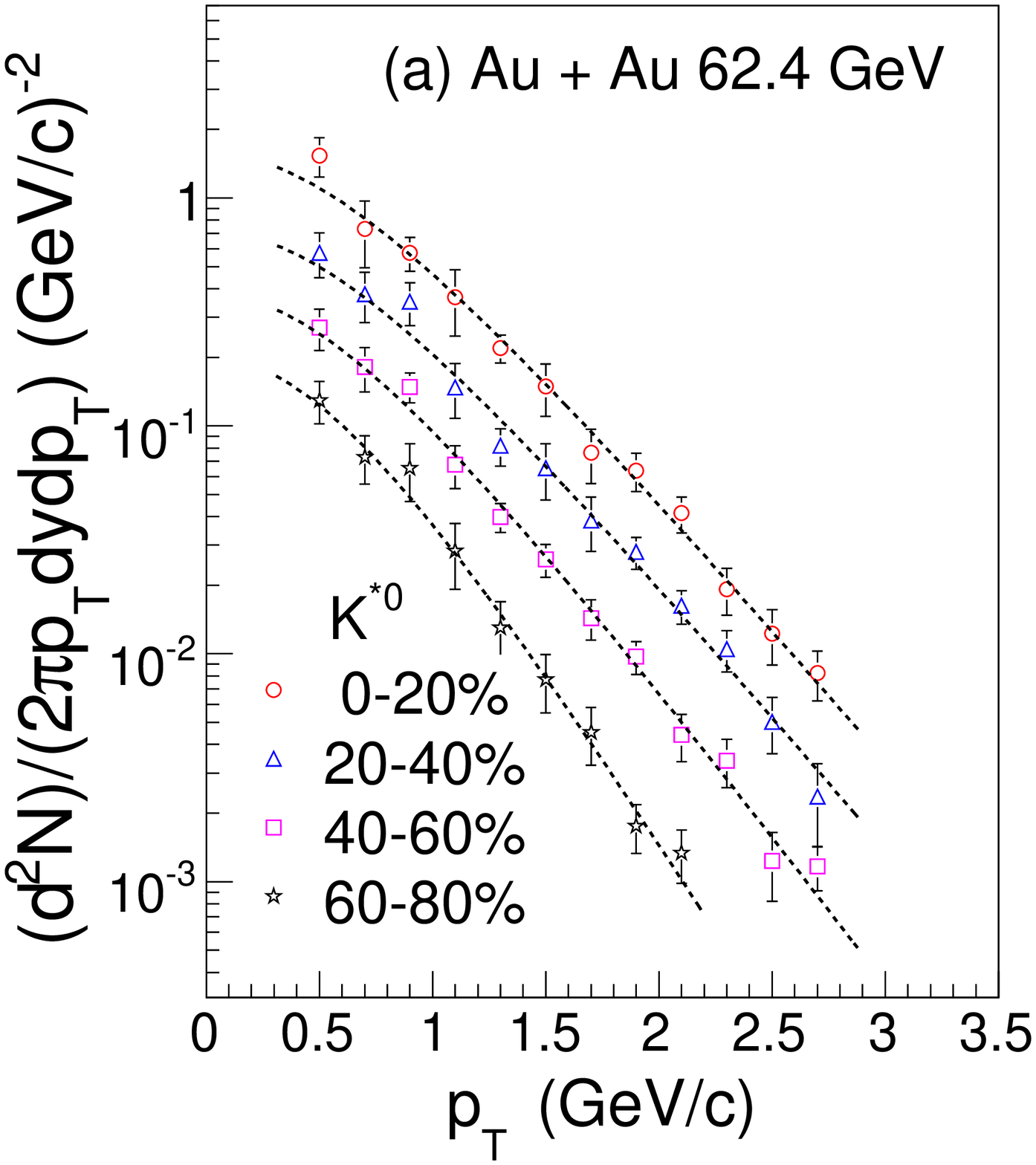}
\includegraphics[scale=0.4]{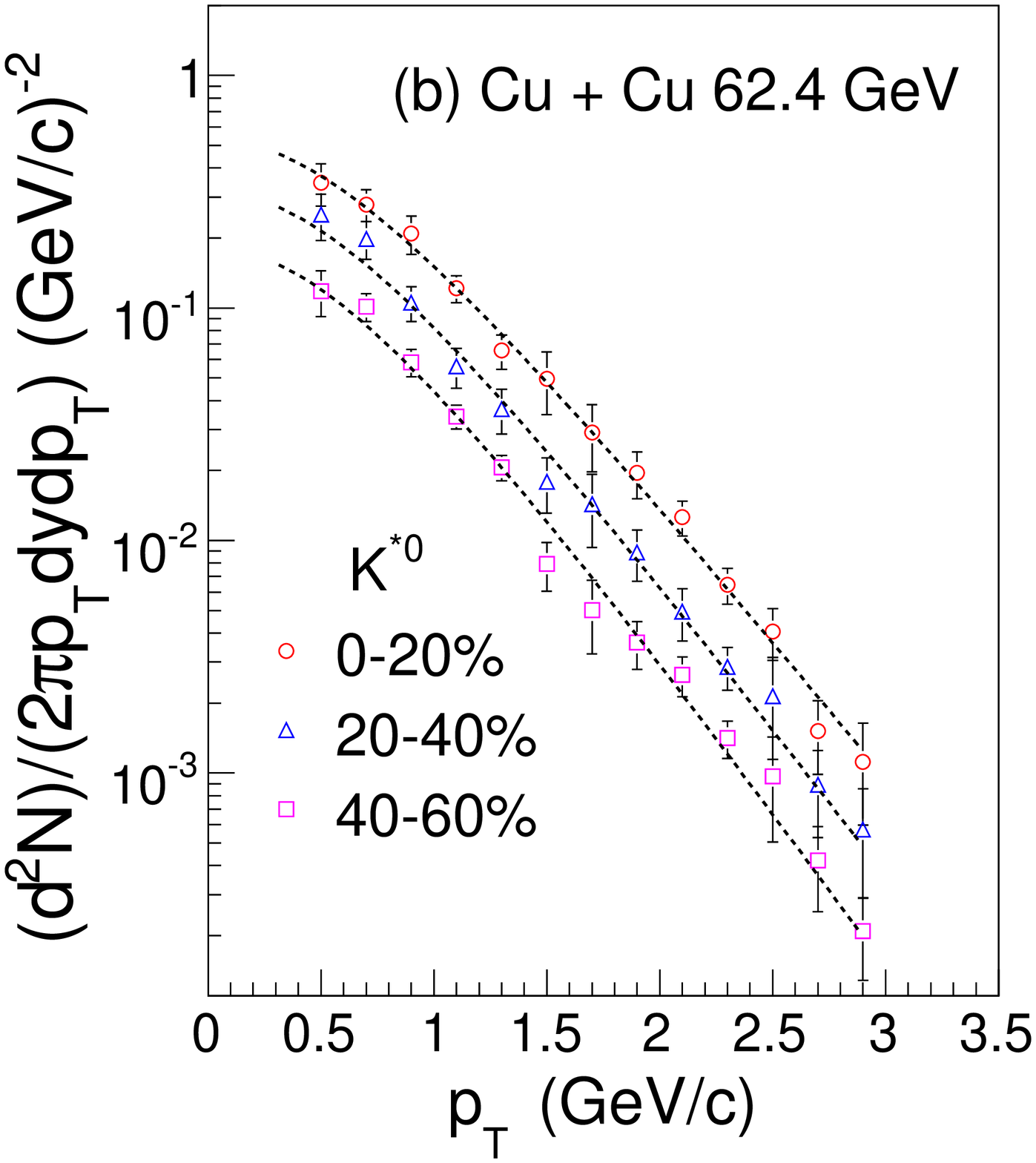}
\includegraphics[scale=0.4]{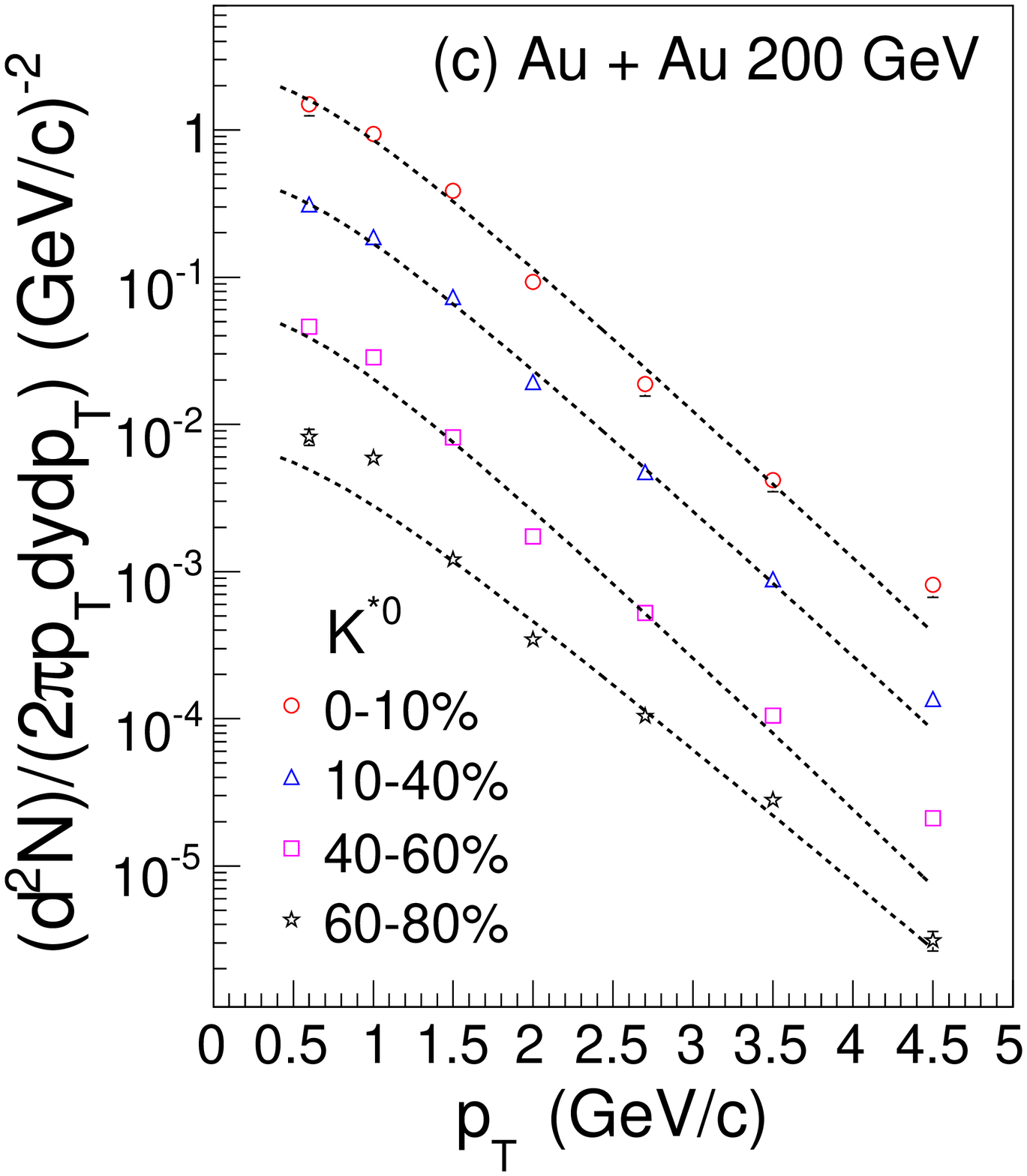}
\includegraphics[scale=0.4]{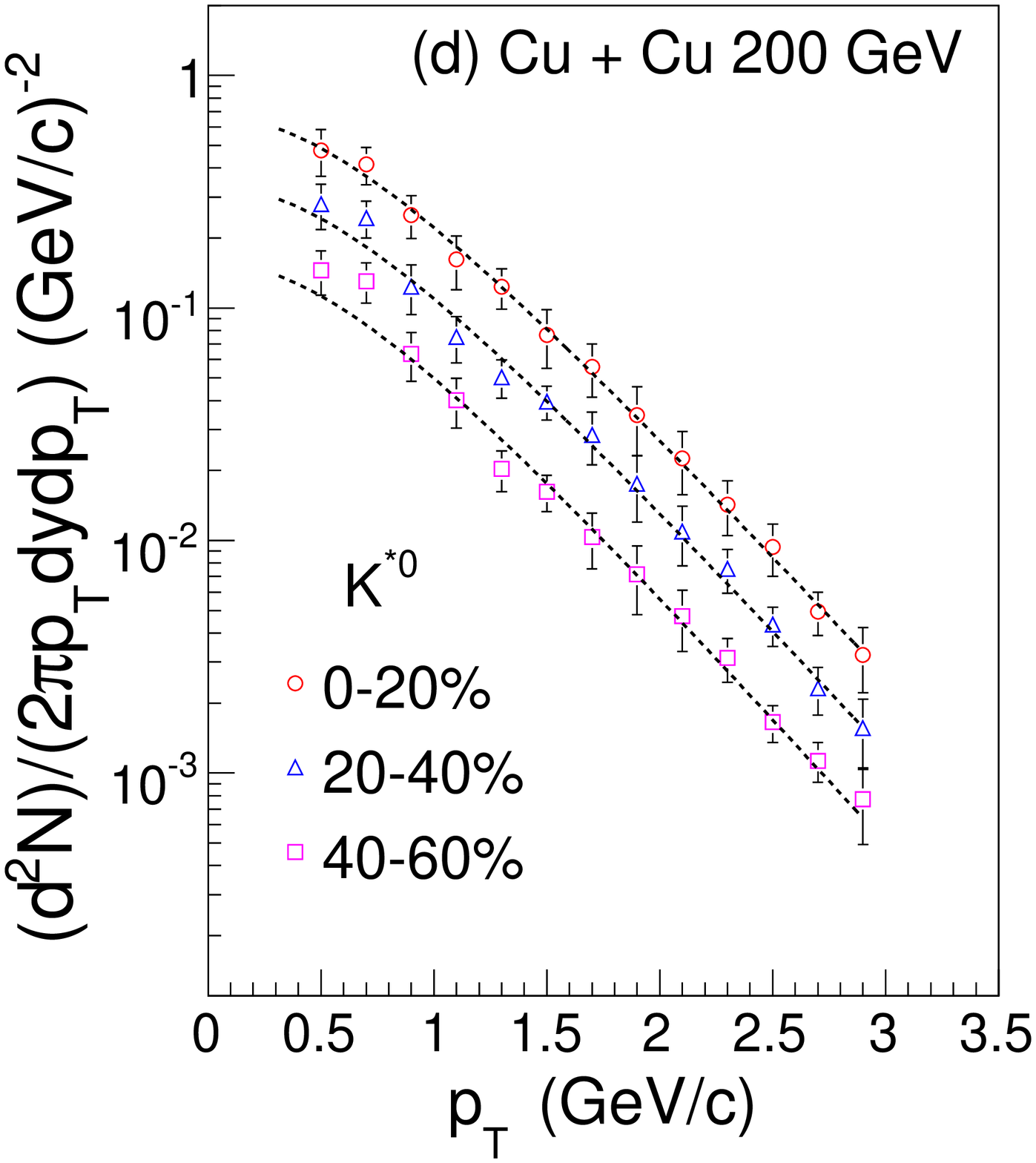}
\caption{Mid-rapidity $K^{*0}$ $p_{\mathrm T}$ spectra for various collision 
centrality bins in Au+Au and Cu+Cu collisions at $\sqrt{s_{\mathrm {NN}}}$ = 62.4 
and 200 GeV. The dashed lines represent the exponential fit to data. The 
errors shown are quadratic sums of statistical and systematic uncertainties.}
\label{spectra}
\end{center}
\end{figure*}
\subsection{Transverse momentum spectra}

The  $K^{*0}$ invariant yields as a function of $p_{\mathrm T}$ were evaluated 
by correcting the extracted raw yields for detector acceptance and reconstruction 
efficiency. The raw yield was obtained by fitting the data to the BW+RBG function. 
The efficiency $\times$ acceptance was obtained by embedding Monte Carlo (MC) 
simulations of kaons and pions from $K^{*0}$ decays into the real data using 
STAR GEANT and passing these embedding data through the same reconstruction chain as for the 
real data \cite{geant}. In addition, the yields were corrected for collision vertex finding 
efficiency and the decay branching ratio of 0.66. The vertex finding efficiency 
 is 94.5\% for Au+Au collisions at 62.4 GeV and 92.2\% for  Cu+Cu collisions 
at 62.4 and 200 GeV. The vertex finding efficiency for Au+Au collisions at 200 GeV is 100\%. 
The variation of efficiency $\times$ acceptance with $p_{\mathrm T}$, for 
various centralities in the Au+Au and Cu+Cu system for $\sqrt{s_{\mathrm {NN}}}$ = 200 GeV, 
is depicted in Figure~\ref{eff}. The absence of centrality dependence in the 
efficiency $\times$ acceptance for the Cu+Cu system is due to small variation in 
total multiplicity across the collision centrality studied compared to 
those for the Au+Au system.

Figure~\ref{spectra} shows the $p_{\mathrm T}$ spectra of $K^{*0}$ at 
mid-rapidity ($|y|<0.5$) in Au+Au and Cu+Cu collisions 
at $\sqrt{s_{\mathrm {NN}}}$ = 62.4  and 200 GeV for different collision 
centralities. The dashed lines are the exponential fits to the  $K^{*0}$ data.
The fitting exponential function is defined as
\begin{equation}
\frac{1}{2\pi m_T}\frac{d^2N}{dydm_T}=\frac{dN/dy}{2\pi T(M_0+T)}e^{-(m_T-M_0)/T}
\end{equation}
where the inverse slope parameter $T$ and yield $dN/dy$ are free parameters. 
$M_{0}$ is the mass of the $K^{*0}$.
The above function is found to provide good fits to the data for both 
collision systems. The $\langle p_{\mathrm T} \rangle$, obtained using the 
above functional form for the  $p_{\mathrm T}$ distributions, are presented in 
the following section together with the mid-rapidity yields $dN/dy$.

\subsection{$dN/dy$ and $\langle p_{\mathrm T} \rangle$}
\begin{figure}[htp]
\begin{center}
\includegraphics[scale=0.43]{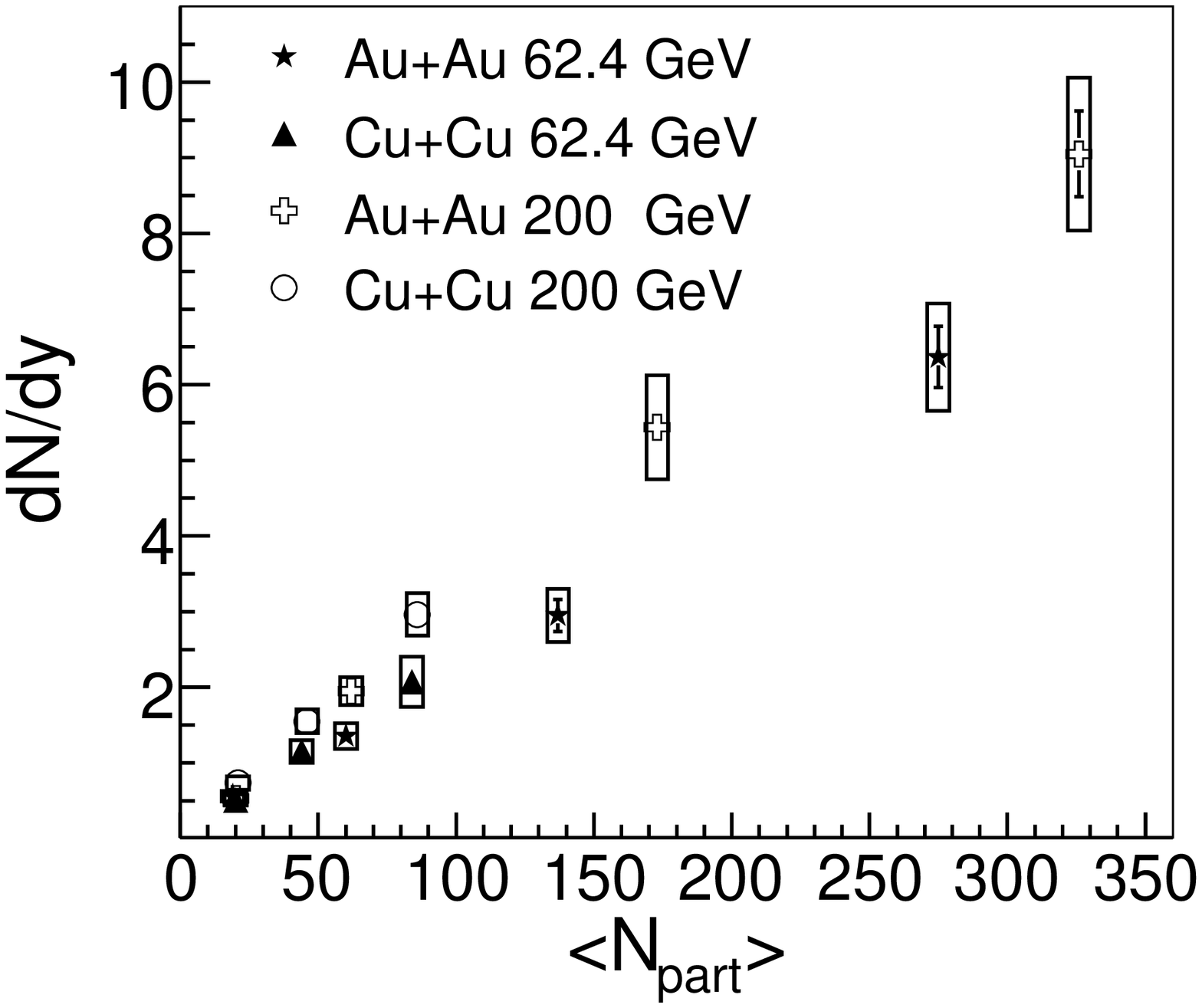}
\caption{The mid-rapidity yields $dN/dy$ of $K^{*0}$ as a function of the 
average number of participating nucleons, $\langle N_{\mathrm {part}} \rangle$, 
for Au+Au and Cu+Cu collisions at $\sqrt{s_{\mathrm {NN}}}$ = 62.4 and 200 GeV. 
The boxes represent the systematic uncertainties.}
\label{dndy}
\end{center}
\end{figure}
The $K^{*0}$ $dN/dy$ yield at mid-rapidity plotted as a function 
of average number of participating nucleons, $\langle N_{\mathrm {part}} \rangle$, is shown 
in Figure~\ref{dndy}. The $dN/dy$ for $K^{*0}$ presented here was calculated 
by using the data points in the measured range of the $p_{\mathrm T}$ spectrum 
while assuming an exponential behavior outside the fiducial range. 
The $K^{*0}$ integrated yield is higher for center of mass energies of 200 
GeV than  62.4 GeV. For collisions at a given beam energy with similar 
$\langle N_{\mathrm {part}} \rangle$, the $dN/dy$ is similar for Au+Au and Cu+Cu systems. 
A similar behavior was observed for $\phi$ mesons at RHIC~\cite{starphiplb}. 

\begin{figure}
\begin{center}
\includegraphics[scale=0.4]{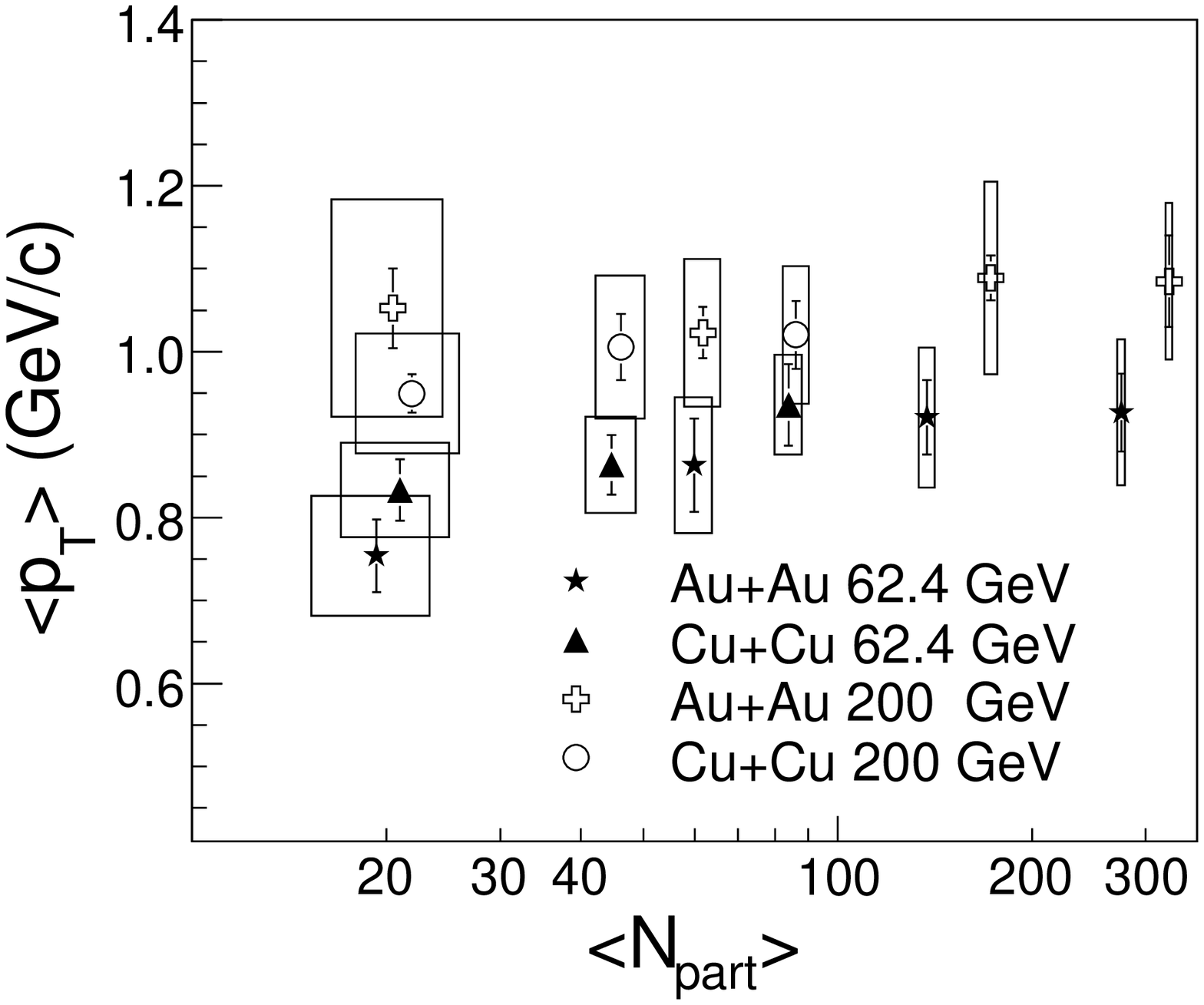}
\caption{The mid-rapidity $K^{*0}$ $\langle p_{\mathrm T} \rangle$ as a 
function $\langle N_{\mathrm {part}} \rangle$ for Au+Au and Cu+Cu collisions 
at $\sqrt{s_{\mathrm {NN}}}$ = 62.4 and 200 GeV. The boxes represent the 
systematic uncertainties.}
\label{meanpt}
\end{center}
\end{figure}

\begin{table*}
\caption{The $K^{*0}$ $dN/dy$ and $\langle p_{\mathrm T} \rangle$  at $|y|< 0.5$ 
measured in Au+Au and Cu+Cu 
collisions at 62.4 GeV and 200 GeV for different collision centralities. Both statistical
and systematic uncertainties are given. The error given for $T$ is statistical only. }
\label{Table:dndy}
\begin{tabular}{cccccc}
\hline
 Collision systems  &  Centrality  & $\langle N_{part} \rangle$ & $dN/dy$  & $T$ (GeV) & $\langle p_{T} \rangle$ (GeV) \\
\hline
Au+Au(62.4 GeV)   & 0-20\%   & 275 & $6.4 \pm 0.4 \pm 0.7$  & $0.36 \pm 0.02$ & $0.93 \pm 0.05 \pm  0.09$ \\
                  & 20-40\% & 137 & $2.9 \pm 0.2 \pm 0.4$ & $0.35 \pm 0.02$ & $0.92 \pm 0.05 \pm  0.08$ \\
                  & 40-60\% & 60 & $1.4 \pm 0.1 \pm 0.2$  & $0.32 \pm 0.01$ & $0.86 \pm 0.06 \pm  0.08$ \\
                  & 60-80\% & 19 & $0.56 \pm 0.03\pm 0.07$ & $0.26 \pm 0.02$& $0.75 \pm 0.04 \pm  0.07$ \\
\hline
Cu+Cu(62.4 GeV)   & 0-20\%  & 84 & $2.07 \pm 0.07 \pm 0.30$ & $0.35 \pm 0.01$ & $0.94 \pm 0.05 \pm 0.06$ \\
                  & 20-40\% & 44 & $1.15 \pm 0.06 \pm 0.20$ & $0.33 \pm 0.01$ & $0.86 \pm 0.04 \pm 0.06$ \\
                  & 40-60\% & 20 & $0.51 \pm 0.03 \pm 0.07$& $0.31 \pm 0.01$ & $0.83 \pm 0.04 \pm 0.06$ \\
\hline
Au+Au(200 GeV)   & 0-10\%  & 326 & $9.05 \pm 0.57 \pm 1.01$ & $0.41 \pm 0.02$ &  $1.09 \pm 0.06 \pm 0.094$ \\
                 & 10-40\% & 173 & $5.43 \pm 0.17 \pm 0.69$ & $0.43 \pm 0.02$ &  $1.09 \pm 0.03 \pm 0.12$ \\
                 & 40-60\% & 62 & $1.95 \pm 0.07 \pm 0.18$ & $0.39 \pm 0.02$ & $1.02 \pm 0.03 \pm0.09$  \\
                 & 60-80\%  & 20 & $0.53 \pm 0.03 \pm 0.08$ & $0.40 \pm 0.03$ & $1.05 \pm 0.05 \pm 0.13$ \\
                  
\hline

Cu+Cu(200 GeV)   & 0-20\%  & 86 & $2.96 \pm 0.12 \pm 0.30$  & $0.40 \pm 0.01$ & $1.0 \pm  0.04 \pm  0.08$ \\
                 & 20-40\% & 46 & $1.55 \pm 0.06 \pm 0.20$  & $0.40 \pm 0.02$ & $1.0 \pm  0.04 \pm  0.09$ \\
                 & 40-60\% & 21 & $0.73 \pm 0.03 \pm 0.09$ & $0.40 \pm 0.02$ & $0.95 \pm 0.02 \pm  0.07$ \\
\hline

p+p (200 GeV)     &          & 2 & $0.005 \pm 0.002 \pm 0.006$ &$0.20 \pm 0.01$  & $0.81 \pm 0.02 \pm 0.14$ \\
\hline        
\hline
\end{tabular}
\end{table*}
\begin{table*}
\caption{The contributions for various sources for estimating the total
systematic uncertainties for
$K^{*0}$ at mid-rapidity ($|y|<$0.5) on $dN/dy$
and $\langle p_{T} \rangle$ in 0-20\% Au+Au collisions at 62.4 GeV.
The systematic uncertainties are similar for other collision systems. }
\label{Table:syst}
\begin{tabular}{ccc}
\hline
{Different Sources}&{$dN/dy$}& {$\langle p_{T} \rangle$ (GeV/$c$)}\\
\hline
exponential fit&0\%&0\%\\
Levy fit&5.24\%&0.44\%\\
Background function &4.49\%&2.88\%\\
(higher order polynomial)&  & \\
Relativistic Breit Wigner &2.64\%&0.542\%\\
$|V_Z|<$ 20 cm&1.6\%&1.05\%\\
Track type ($K^{*0}$)&3.6\%&4.3\%\\
Track type ($\overline K^{*0}$ )&4.05\%&5.96\%\\
NFitPnts = 22&4.32\%&1.35\%\\
$|N_{\sigma\pi}, N_{\sigma K}| < 3$&4.45\%&4.87\%\\
\hline
Total Sys. Uncertainty& 11.18\% &9.47\%\\
\hline
\end{tabular}
\end{table*}

The $K^{*0}$ $\langle p_{\mathrm T} \rangle$ at mid-rapidity plotted as a function 
of $\langle N_{\mathrm {part}} \rangle$, is shown in Figure~\ref{meanpt} for 
Au+Au and Cu+Cu collisions at $\sqrt{s_{\mathrm {NN}}}$ = 62.4 and 200 GeV.
The $\langle p_{\mathrm T} \rangle$ for $K^{*0}$ presented here was calculated 
by using the data points in the measured range of the $p_{\mathrm T}$ spectrum 
while assuming an exponential behavior outside the fiducial range. 
No significant centrality and colliding ion size dependence could be observed.
However, the  $\langle p_{\mathrm T} \rangle$ values 
for collisions at 200 GeV are seen to be slightly higher 
than those from 62.4 GeV in both Au+Au and Cu+Cu collisions.
 Previous measurements of $\langle p_{\mathrm T} \rangle$ of $K^{*0}$ 
in heavy ion collisions have been shown to be higher than the
 corresponding values in $p$+$p$ collisions~\cite{haibinPRC}. 
This may be understood from the following.
According to UrQMD transport model calculations~\cite{urqmd}, $K^{*0}$s are
more likely to be reconstructable in the high $p_{\mathrm T}$  
region than in the low $p_{\mathrm T}$ region. This is because  high 
$p_{\mathrm T}$ $K^{*0}$s are more likely to escape the medium before 
the kinetic freeze-out stage and are thus less affected by in-medium 
effects~\cite{thesis,sadhana}. 

\begin{figure}
\begin{center}
\includegraphics[scale=0.4]{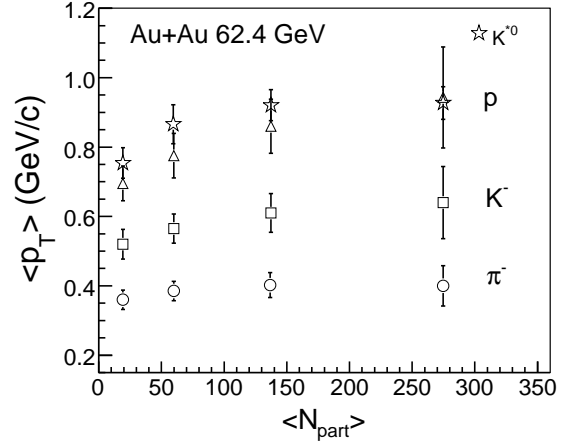}
\caption{The mid-rapidity $\langle p_{\mathrm T} \rangle$ of $\pi$, K, $p$ 
and $K^{*0}$ as a function of $\langle N_{\mathrm {part}} \rangle$ for Au+Au 
collisions at $\sqrt{s_{\mathrm {NN}}}$ = 62.4 GeV.}
\label{meanpt1}
\end{center}
\end{figure}

Figure~\ref{meanpt1} shows the $\langle p_{\mathrm T} \rangle$ of different 
particle species ($\pi$, $K$, $p$, and $K^{*0}$) in Au+Au collision at 
$\sqrt{s_{\mathrm {NN}}}$ = 62.4 GeV as a function of  
$\langle N_{\mathrm {part}} \rangle$. The 
$\langle p_{\mathrm T} \rangle$ of $K^{*0}$ is higher 
than the $\langle p_{\mathrm T} \rangle$ of kaons and pions and closer to 
that of protons. This indicates that 
the $\langle p_{\mathrm T} \rangle$ is strongly coupled with the mass of the 
particle, in agreement with similar observations made previously in Au+Au and
d+Au collisions at 200 GeV \cite{haibinPRC,dAu}. 
Table~\ref{Table:dndy} lists the $dN/dy$ and $\langle p_{\mathrm T} \rangle$ of 
$K^{*0}$ for various collision systems at different collision centralities 
and beam energies studied.

The systematic uncertainties on $K^{*0}$ $dN/dy$ and $\langle p_{\mathrm T} 
\rangle$ were estimated as follows \cite{sadhana}:
(1) an uncertainty on the $K^{*0}$ signal fit of the invariant mass spectrum was 
evaluated by replacing the non-relativistic BW function with a relativistic 
BW function, (2) an uncertainty on the background distribution fit was evaluated by using 
a higher order polynomial function, (3) by varying the track types  
($K^{*0}$ and $\overline{K^{*0}}$), (4) using different functions 
such as a Levy function \cite{levy,starpiplb} 
to fit the spectra, (5) an uncertainty on track selection was estimated by varying the 
track cuts such as $N_{\sigma}$ cut, NFitPnts cut, and (6) by varying 
the $V_{Z}$ cut from 30 cm to 20 cm.
The systematic uncertainties coming from the different sources are listed in 
Table ~\ref{Table:syst} for Au+Au collisions at 62.4 GeV.
\begin{figure}[htp]
\begin{center}
\includegraphics[scale=0.35]{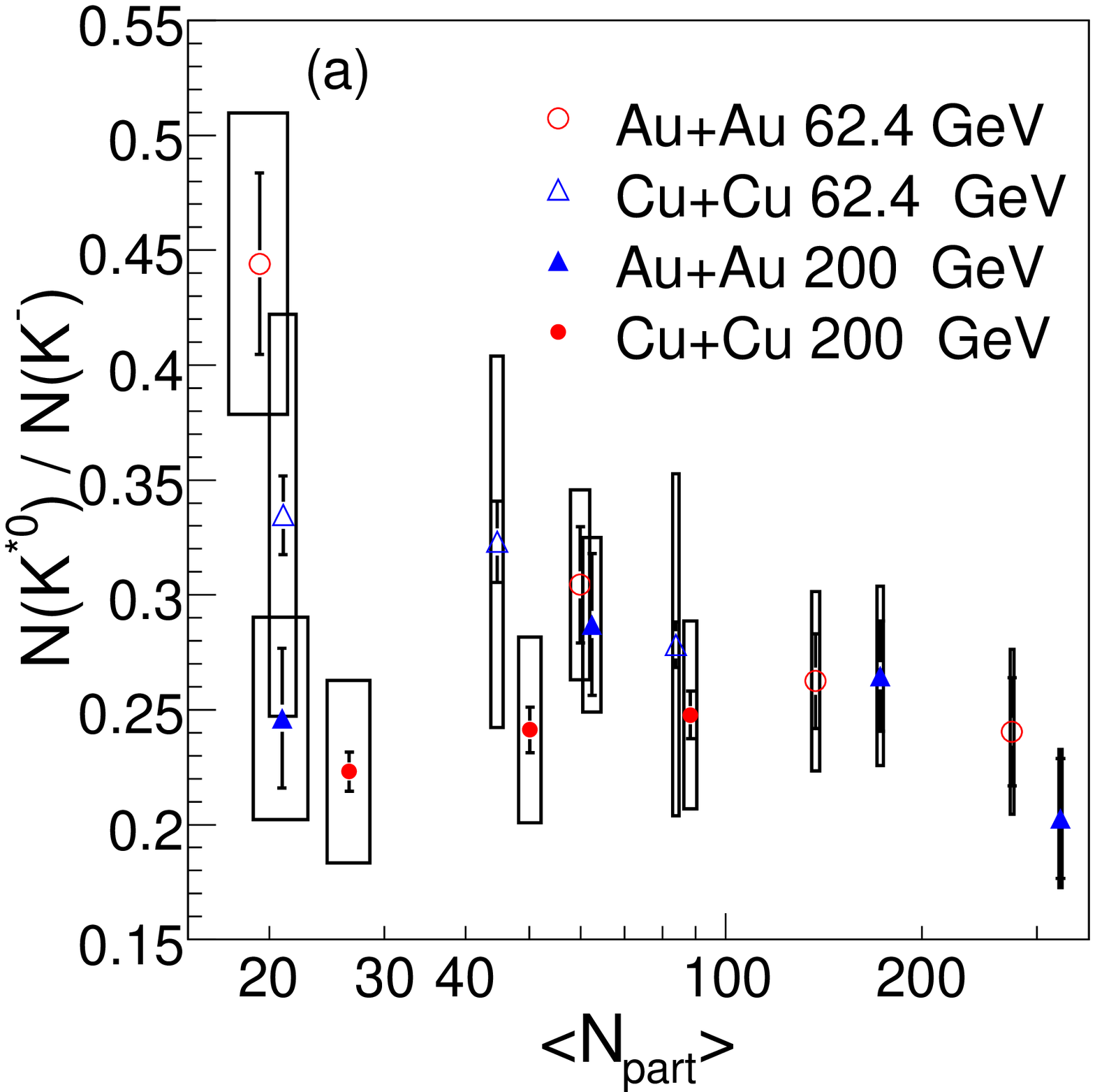}
\includegraphics[scale=0.35]{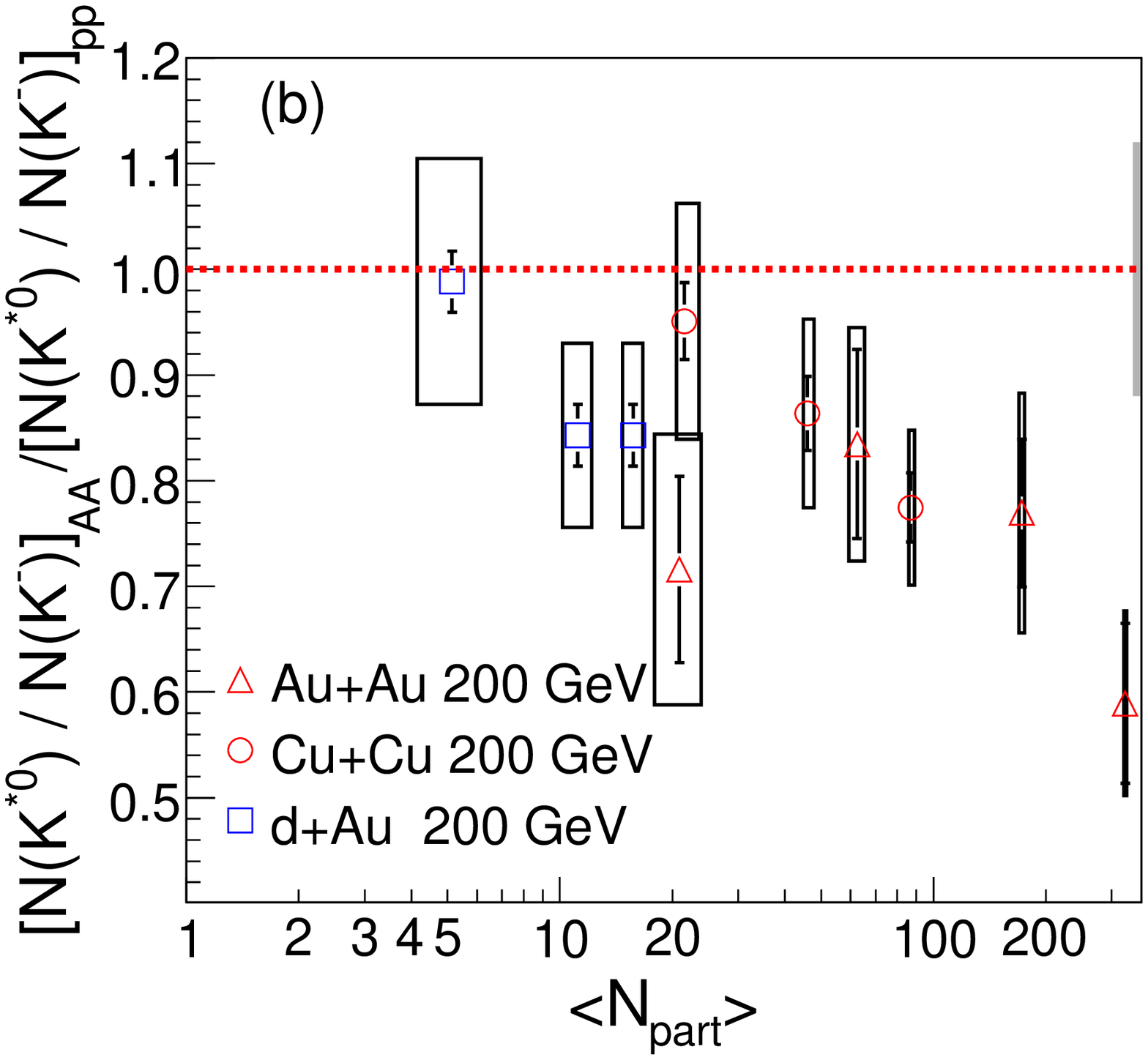}
\includegraphics[scale=0.35]{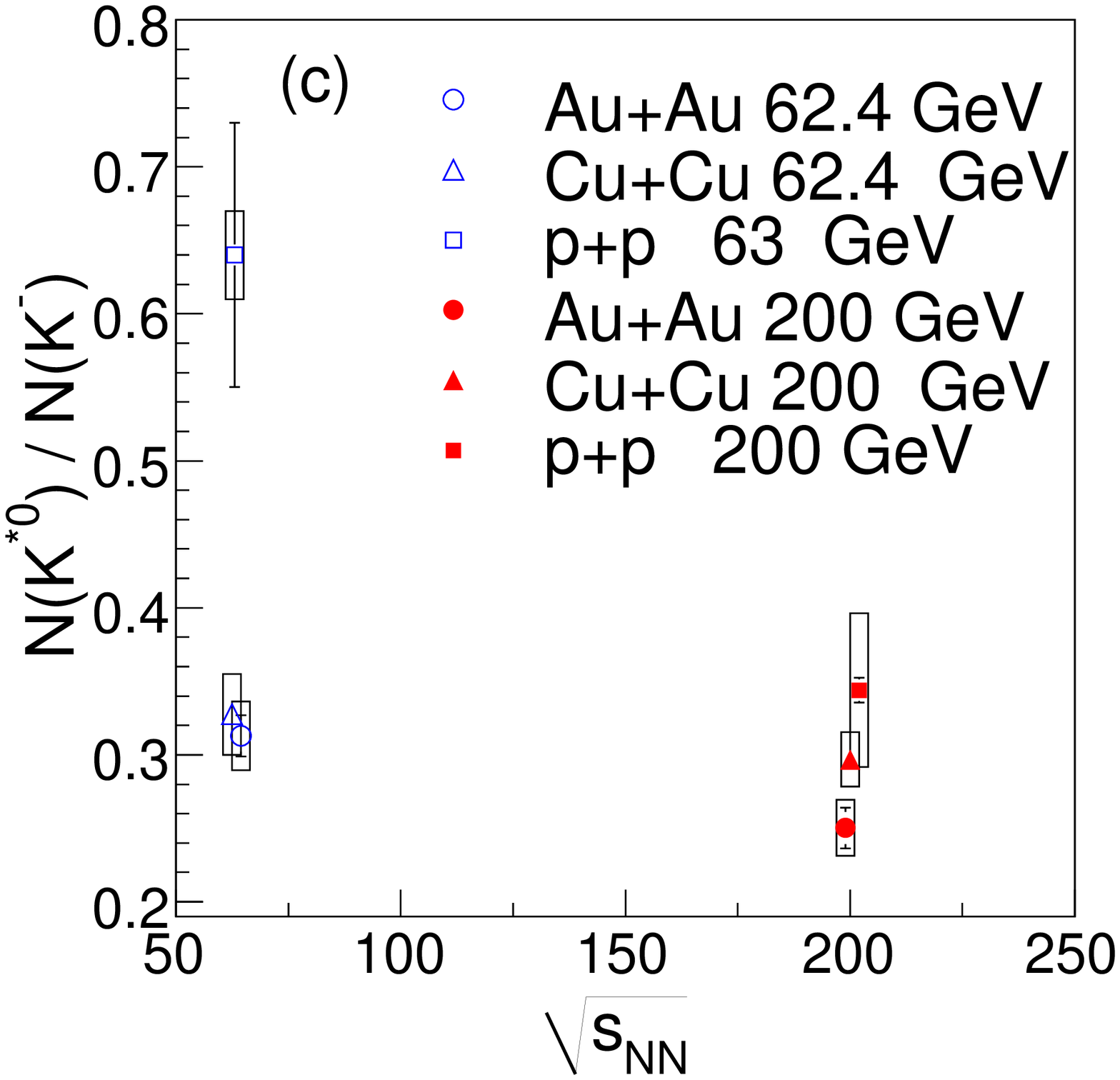}
\caption{(a) Mid-rapidity N($K^{*0}$)/N($K^{-}$) ratio for Au+Au and Cu+Cu collisions at 
$\sqrt{s_{\mathrm {NN}}}$ = 62.4 and 200 GeV as a 
function of $\langle N_{\mathrm {part}} \rangle$.
(b) Mid-rapidity $N(K^{*0})/N(K^{-})$ in Au+Au, Cu+Cu and d+Au collisions 
divided  by $N(K^{*0})/N(K^{-})$ ratio in $p$+$p$ collisions  at  
$\sqrt{s_{\mathrm {NN}}}$ = 200 GeV 
as a function of $\langle N_{\mathrm {part}} \rangle$. 
(c) Mid-rapidity  $N(K^{*0})/N(K^{-})$ ratio 
in minimum bias Au+Au, Cu+Cu, $p$+$p$ 
collisions as a function of $\sqrt{s_{\mathrm {NN}}}$. The boxes represents
systematic uncertainties. The value of $N(K^{*0})/N(K^{-})$
 ratio in $p$+$p$ at 63 GeV is from ISR~\cite{isr}.}
\label{particleratio}
\end{center}
\end{figure}
\begin{figure}[htp]    
\begin{center}
\includegraphics[scale=0.35]{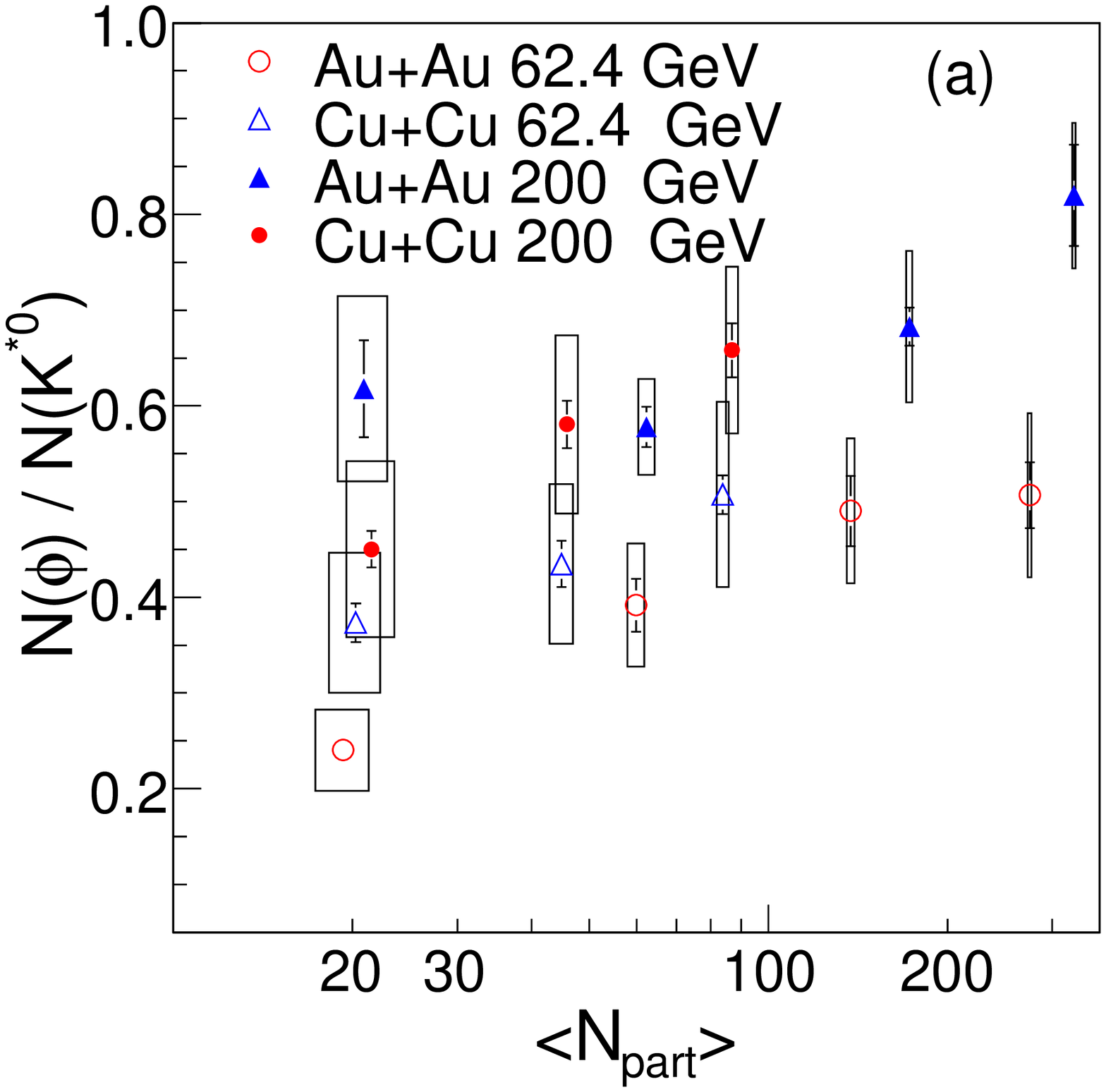}
\includegraphics[scale=0.35]{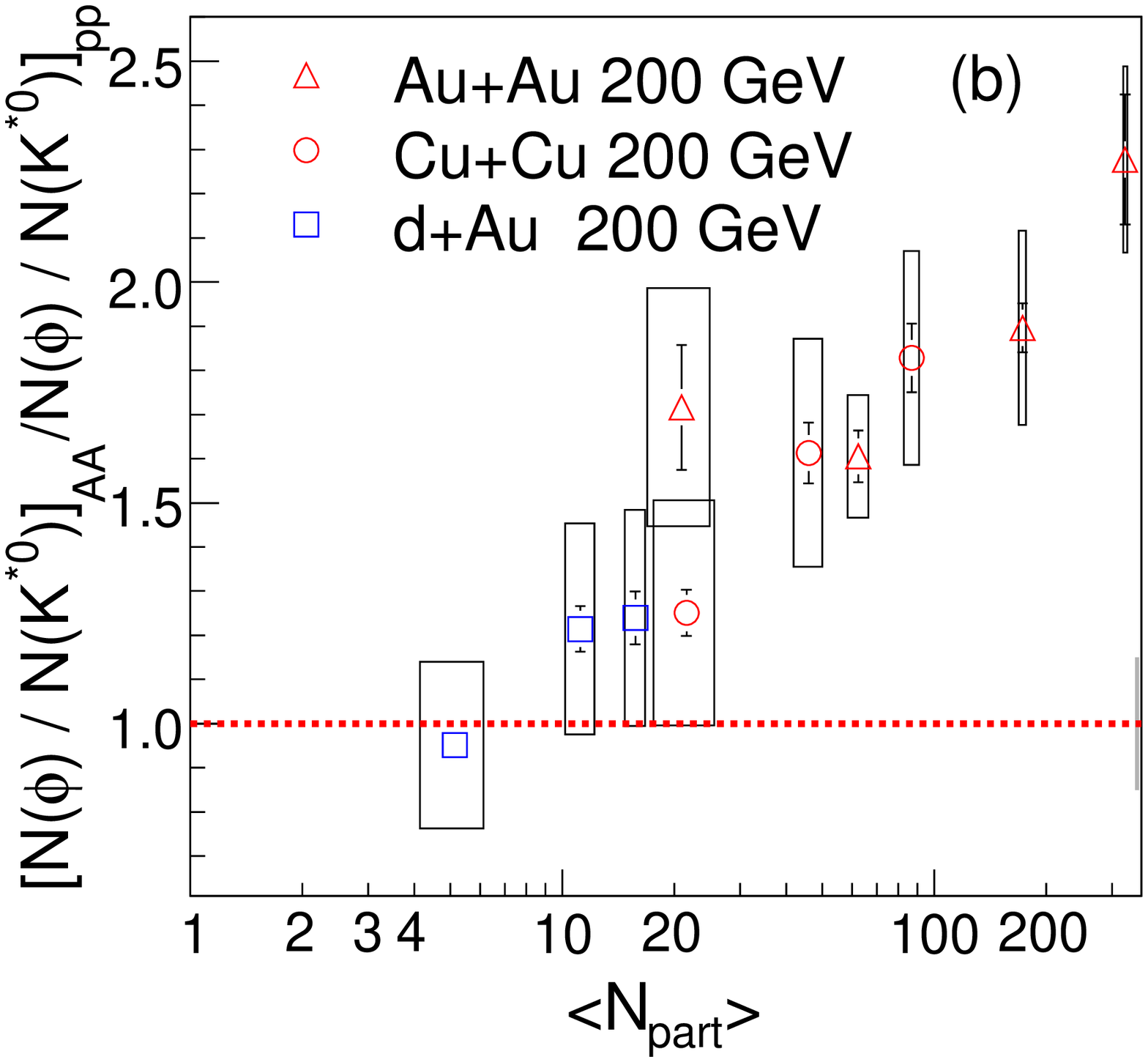}
\includegraphics[scale=0.35]{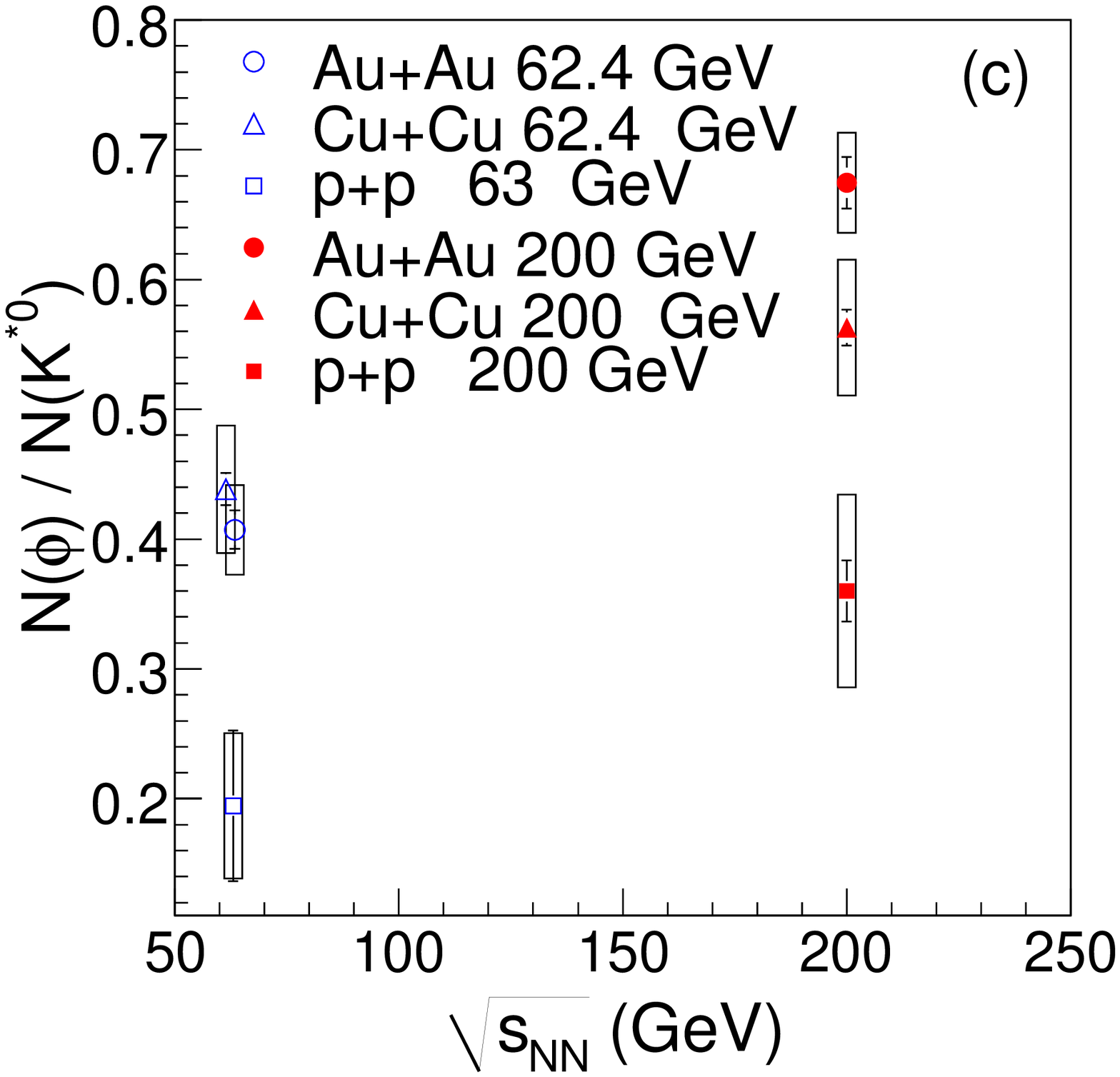}
\caption{(a) Mid-rapidity $N(\phi)/N(K^{*0})$ ratio for Au+Au and Cu+Cu collisions at 
$\sqrt{s_{\mathrm {NN}}}$ = 62.4 and 200 GeV as a 
function of $\langle N_{\mathrm {part}} \rangle$.
(b) Mid-rapidity $N(\phi)/N(K^{*0})$ in Au+Au, Cu+Cu and d+Au collisions 
divided  by $N(\phi)/N(K^{*0})$
ratio in $p$+$p$ collisions 
at  $\sqrt{s_{\mathrm {NN}}}$ = 200 GeV 
as a function of $\langle N_{\mathrm {part}} \rangle$. 
(c) Mid-rapidity $N(\phi)/N(K^{*0})$ ratio 
in minimum bias Au+Au, Cu+Cu, $p$+$p$ collisions as a 
function of $\sqrt{s_{\mathrm {NN}}}$. The boxes represents
systematic uncertainties. The values of $N(\phi)/N(K^{*0})$ 
ratio in $p$+$p$ at 63 GeV is from ISR~\cite{isr}.}
\label{particleratio1}
\end{center}
\end{figure}

\subsection{Particle Ratio}

Figure~\ref{particleratio}(a) shows the ratio of $K^{*0}$ and $K^{-}$ 
yields, $N(K^{*0})/N(K^{-})$ as 
a function of $\langle N_{\mathrm {part}} \rangle$ in Au+Au and Cu+Cu collisions 
at $\sqrt{s_{\mathrm {NN}}}$ = 62.4  and 200 GeV. From the figure, no clear beam
energy or system size dependence is observed. The Fig.\ref{particleratio}(b)
shows the $K^{*0}/K^{-}$ ratio in Au+Au, Cu+Cu, and $d$+Au collisions normalized 
by their corresponding values measured in $p$+$p$ collisions at 
$\sqrt{s_{\mathrm {NN}}}$ = 200 GeV. This $N(K^{*0})/N(K^{-})$ double ratio 
is seen to be much smaller than unity in central Au+Au collisions. In contrast the 
$N(K^{*0})/N(K^{-})$ double ratio is close to unity for $d$+Au collisions. 
This suggests strong re-scattering of decay daughters of $K^{*0}$ meson, 
resulting in the loss of reconstructed $K^{*0}$ signal. The re-scattering of $K^{*0}$ 
daughter particles depends on $\sigma_{\pi\pi}$ which is considerably larger 
than $\sigma_{\pi K}$, but $\sigma_{\pi K}$ is responsible for regeneration of the
$K^{*0}$ meson. Therefore, we expect a decrease of the $N(K^{*0})/N(K^{-})$ 
yield ratio in heavy 
ion collisions owing to strong re-scattering of  $K^{*0}$ daughter particles. The 
observed decrease in the $K^{*0}/K^{-}$ double ratio indicates an extended 
lifetime for the hadronic phase as we move from  p+p and  d+Au to  Au+Au 
collisions. The extended lifetime enhances the re-scattering effect.
Figure~\ref{particleratio}(c) shows the energy dependence of the
$N(K^{*0})/N(K^{-})$ ratio for minimum bias Au+Au and Cu+Cu collisions at 
$\sqrt{s_{\mathrm {NN}}}$ = 62.4  and 200 GeV. Also included in the figure 
are values obtained from $p$+$p$ collisions at 63 GeV ~\cite{isr} and 200 
GeV~\cite{haibinPRC}. At both energies, the $N(K^{*0})/N(K^{-})$ for $p$+$p$ 
collisions is higher than the values in the heavy ion 
collisions. This can be attributed to larger re-scattering of $K^{*0}$ 
daughter particles in heavy ion collisions.

Another ratio of considerable interest is the $N(\phi)/N(K^{*0})$ ratio as both 
the $\phi$
and $K^{*0}$ have the same spin and similar mass, but different strangeness  
and lifetime. The lifetime of the $\phi$ meson is 40 fm/$c$ ($\sim$10 times that of 
$K^{*0}$). Due to the relatively longer lifetime of the $\phi$ meson and negligible 
$\sigma_{KK}$~\cite{starphiplb}, we expect both the re-scattering and regeneration 
effects to be negligible for the $\phi$ meson. Since $\phi$ has two strange 
quarks and $K^{*0}$ has one, $N(\phi)/N(K^{*0})$ can also give information regarding 
strangeness enhancement.

\begin{table*}
\caption{The  mid-rapidity $N(K^{*0})/N(K^{-})$  and $N(\phi)/N(K^{*0})$ 
yield ratio in Au+Au and Cu+Cu collisions
at $\sqrt{s_{\mathrm {NN}}}$ = 62.4 GeV and 200 GeV for different centralities.
The first uncertainty is statistical and the second one is systematic.}
\label{Table:ratio}
\begin{tabular}{cccc}
\hline
 Collision systems  &  Centrality  & $N(K^{*0})/N(K^{-})$  & $N(\phi)/N(K^{*0})$ \\
\hline
Au+Au(62.4 GeV)   & 0-20\%  & $0.24 \pm 0.02 \pm 0.04$  &  $0.51 \pm 0.03 \pm 0.09$ \\
                  & 20-40\% & $0.26 \pm 0.02 \pm 0.04$  &  $0.49 \pm 0.03 \pm 0.08$ \\
                  & 40-60\% & $0.30 \pm 0.03 \pm 0.04$  &  $0.39 \pm 0.03 \pm 0.06$ \\
                  & 60-80\% & $0.44 \pm 0.04 \pm 0.07$  &  $0.24 \pm 0.02 \pm 0.04$ \\
\hline
Cu+Cu(62.4 GeV)   & 0-20\%  &  $0.29 \pm 0.01 \pm 0.05$  &  $0.51 \pm 0.02 \pm 0.1$  \\
                  & 20-40\% &  $0.34 \pm 0.02 \pm 0.04$  &  $0.43 \pm 0.02 \pm 0.08$ \\
                  & 40-60\% &  $0.36 \pm 0.02 \pm 0.05$  &  $0.37 \pm 0.02 \pm 0.07$\\
\hline

Au+Au(200 GeV)    & 0-10\%  &  $0.20  \pm 0.03 \pm 0.03$  & $0.82 \pm 0.05 \pm 0.08$ \\
                  & 10-40\% & $0.26   \pm 0.02 \pm 0.04$  & $0.68 \pm 0.02 \pm 0.08$\\
                  & 40-60\% &  $0.29  \pm 0.03 \pm 0.04$  & $0.58 \pm 0.02 \pm 0.05$ \\
                  & 60-80\% &   $0.25 \pm 0.03 \pm 0.04$  & $0.62 \pm 0.05 \pm 0.1$\\
\hline

Cu+Cu(200 GeV)    & 0-20\%  &  $0.27 \pm 0.01 \pm 0.03$ & $0.66  \pm 0.03 \pm 0.09$ \\
                  & 20-40\% & $0.30  \pm 0.01 \pm 0.03$ & $0.58  \pm 0.02 \pm 0.09$ \\
                  & 40-60\% & $0.33  \pm 0.01 \pm 0.04$  & $0.45 \pm 0.02 \pm 0.09$ \\

\hline
p+p(200 GeV)    & MB  &  $0.34 \pm 0.01 \pm 0.05$ & $0.36  \pm 0.02 \pm 0.07$ \\
\hline
\end{tabular}
\end{table*}

Figure~\ref{particleratio1}(a) depicts the $N(\phi)/N(K^{*0})$ ratio as a function
of $\langle N_{\mathrm {part}} \rangle$, corresponding to Au+Au and Cu+Cu 
collisions at $\sqrt{s_{\mathrm {NN}}}$ = 62.4 and 200 GeV. We observe that
the ratio tends to increase with increasing $\langle N_{\mathrm {part}} \rangle$ 
at a given beam energy. The $N(\phi)/N(K^{*0})$ ratio is  higher for 
$\sqrt{s_{\mathrm {NN}}}$ = 200 GeV compared to $\sqrt{s_{\mathrm {NN}}}$ = 
62.4 GeV for the various collision centralities. At a given beam energy and 
$\langle N_{\mathrm {part}} \rangle$ the $N(\phi)/N(K^{*0})$ 
ratio is similar for Au+Au and Cu+Cu collisions. Figure~\ref{particleratio1}(b) 
also shows $N(\phi)/N(K^{*0})$ ratio in Au+Au, Cu+Cu, and $d$+Au collisions 
normalized by their corresponding values measured in $p$+$p$ collisions 
at $\sqrt{s_{\mathrm {NN}}}$ = 200 GeV. We observe that this double ratio 
increases with collision centrality, which favors the re-scattering scenario of $K^{*0}$ 
daughter particles. The observed increase can also have contributions from 
strangeness enhancement in more central collisions~\cite{starphiplb}.
Figure~\ref{particleratio1}(c) shows the energy dependence of the 
$N(\phi)/N(K^{*0})$ ratio for minimum bias Au+Au and Cu+Cu collisions 
at $\sqrt{s_{\mathrm {NN}}}$ = 62.4 and 200 GeV and its values from $p$+$p$ 
collisions. The $\sqrt{s}$ = 63 GeV value for $p$+$p$ collisions is from ISR 
measurements~\cite{isr}. 
At both the energies the $N(\phi)/N(K^{*0})$
ratios for $p$+$p$ collisions are lower than the corresponding values 
in  Au+Au collisions. Furthermore, 
there is an indication of an increase of the value with beam energy.
The study of both ratios  $N(K^{*0})/N(K^{-})$ and  $N(\phi)/N(K^{*0})$, as a 
function of colliding species, collision centrality, and beam energy favors the 
re-scattering  scenario over $K^{*0}$ regeneration. The values of the
ratios along with the associated uncertainties are shown in Table~\ref{Table:ratio}.

\subsection{Elliptic Flow}

We apply the standard reaction plane method as employed in Refs \cite{v2a,v2b} 
for the analysis of elliptic flow.
Here, for a given $p_T$ window, the second
order reaction plane angle, $\psi_2$, was determined event-by-event. 
We have different event planes for every $K^{*0}$ candidate, because for 
every $K^{*0}$ candidate, its daughter particles are excluded from the 
event plane determination.
This was later subtracted from the azimuthal angle $\phi$ of each track
in the same event to generate an event plane subtracted azimuthal 
distribution in the variable $\Phi=(\phi-\psi_2)$. The corresponding 
distribution, $d^2N/{dp_Td\Phi}$, in the azimuthal angle $\Phi$ for 
all the events in a given $p_T$ bin were then fitted with a function 
$A[1+2v_2^{obs}\cos(2\Phi)]$ 
where $A$ is a constant. The fitted value of $v_2^{obs}$ was then divided by 
the reaction plane resolution factor to obtain $v_2$ for the $p_{\mathrm T}$ 
window considered \cite{v2b}.

\begin{figure}
\begin{center}
\includegraphics[scale=0.4]{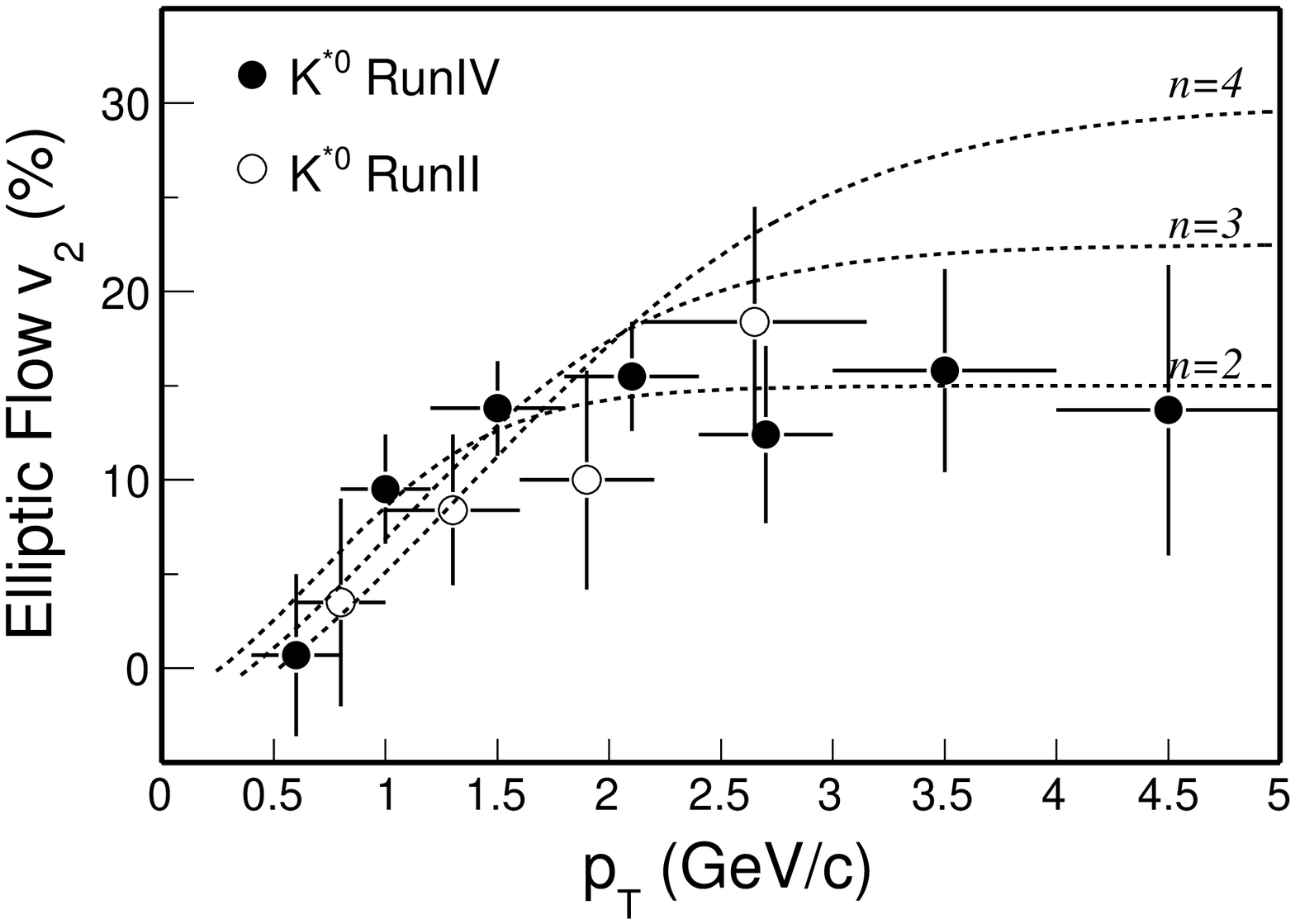}
\caption{The $K^{*0}$ $v_2$ as a function of $p_{\mathrm T}$ in minimum 
bias Au+Au collisions at $\sqrt{s_{\mathrm{NN}}}$ = 200 GeV. Only 
statistical uncertainties are shown. The dashed lines represent the $v_2$ of hadrons
with different number of constituent quarks.}
\label{v2}
\end{center}
\end{figure}

Figure~\ref{v2} shows $v_2$ of $K^{*0}$ as a function of $p_T$ in minimum 
bias Au+Au collisions at $\sqrt{s_{\mathrm{NN}}}$ = 200 GeV. We fit the data 
using the function 
\begin{equation}\label{v2func}
v_2(p_T,n) = \frac{an}{1+exp(-(p_T/n-b)/c)}-dn
\end{equation}
where $a,b,c$, and $d$ are the parameters extracted from such a fit to $v_2$
data obtained earlier for $K^{0}_S$ and $\Lambda$~\cite{Xinv2}. Here $n$, the 
number of constituent quarks, is the only free parameter. The best
fit of the $K^{*0}$ data with the function as given in Eqn.\ref{v2func} yields 
a value of $n=2.0\pm0.4$ ($\chi^2/ndf$ = 2/6). 
A similar fit of the combined results of Run II and Run IV data, taken 
in the years 2002 and 2004, respectively, also yields an identical 
value of $n=2.0\pm0.4$ ($\chi^2/ndf=4/10$). This indicates that $K^{*0}$ 
are dominantly produced from direct quark combinations, and the regenerated 
$K^{*0}$ component in the hadronic stage is negligible compared to 
the primordial $K^{*0}$.

\subsection{Nuclear Modification Factor}

Through a measurement of the nuclear modification factors $R_{CP}$ and $R_{AA}$, 
one probes the dynamics of particle production during hadronization and in-medium
effects~\cite{starpiplb,starpiprl}.
The nuclear modification factor $R_{CP}$, which is the ratio of the invariant 
yields for central to peripheral collisions, normalized by number of binary 
collisions, $N_{bin}$, is defined as
\begin{equation}
R_{CP}=\frac{[dN/(N_{bin}dp_{T})]^{central}}{[dN/(N_{bin}dp_{T})]^{peripheral}}
\end{equation}
where $N_{bin}$ is calculated from the Glauber model~\cite{spectralongpid}. We expect 
$R_{CP}$ to be unity at high $p_{\mathrm T}$ ($>$ 2 GeV/$c$) 
if nucleus-nucleus collisions were mere 
superpositions of nucleon-nucleon collisions. Any deviation observed from unity 
would indicate the presence of in-medium effects. Above 
$p_{\mathrm T}$ = 2 GeV/$c$, the $R_{CP}$ of $\pi^{\pm}$, $p+\bar{p}$, $K^{0}_{S}$, and 
$\Lambda$, as measured by STAR, are found to be significantly lower than 
unity. This suggests a suppression of particle production at high $p_{\mathrm T}$ 
in central collisions relative to peripheral ones~\cite{starpiplb,rcp2,starpiprl}.
Theoretically, this is attributed to the energy loss of highly energetic partons 
while traversing through the dense medium created in heavy ion collisions. We 
also observe that the $R_{CP}$ of $K^{0}_{S}$ and $\Lambda$ are different. Since 
the mass of $K^{*0}$ is close to that of baryons such as $p$ and $\Lambda$, 
a comparison of $R_{CP}$ of $K^{*0}$ with those for $K^{0}_{S}$ and $\Lambda$ can 
be used to understand whether the observed differences in the $R_{CP}$ of the 
$K^{0}_{S}$ and the $\Lambda$ are tied to the particle mass or the baryon-meson 
effect~\cite{rcp2}.

\begin{figure}
\begin{center}
\includegraphics[scale=0.4]{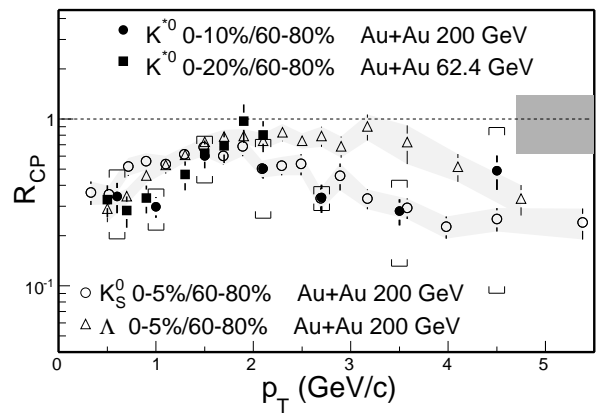}
\caption{The $K^{*0}$ $R_{CP}$ as a function of $p_T$ in Au+Au
collisions at 200 GeV and 62.4 GeV compared to the $R_{CP}$ of
$K^{0}_S$ and $\Lambda$ at 200 GeV. The brackets around Au+Au 200 GeV 
data points are the systematc errors.}
\label{rcp}
\end{center}
\end{figure}

Figure~\ref{rcp} shows the $K^{*0}$ $R_{CP}$ as a function of $p_{\mathrm T}$ 
compared to those for $\Lambda$ and $K^{0}_S$ ~\cite{rcp2}. The shaded band 
around the data points represents the systematic uncertainties and the band around 1
on the right corner represents the normalization uncertainty. For Au+Au collisions 
at 200 GeV the $K^{*0}$ $R_{CP}$ was obtained from the $p_{\mathrm T}$ spectra 
of top 10\% and 60-80\% centrality classes. For Au+Au collisions at 62.4 GeV 
the $p_T$ spectra of the top 20\% and 60-80\% centrality classes were considered. 
The $\Lambda$ and $K^{0}_S$ $R_{CP}$ correspond to the $p_T$ spectra of the 
top 5\% and 60-80\% Au+Au collisions 
at 200 GeV \cite{rcp2}. For $p_T <$ 1.8 GeV/c, the $R_{CP}$ of $K^{*0}$ in Au+Au 
collisions at 200 GeV and 62.4 GeV are smaller than that of $\Lambda$ and 
$K^{0}_S$. This is consistent with the assumption that the re-scattering effect 
dominates over the regeneration effect for $K^{*0}$ at low $p_{\mathrm T}$. 
For Au+Au collisions at
200 GeV, for $p_{\mathrm T} >$ 1.8 GeV/c, the $R_{CP}$ of $K^{*0}$ is closer to 
that for $K^{0}_S$  (differing from that of  $\Lambda$). 
Since the masses of the $\Lambda$ and $K^{*0}$ are similar, the observed difference 
seems to be due to other than mass. The observed 
differences might arise because the $\Lambda$ is a baryon 
whereas $K^{*0}$ is a meson. This 
supports the quark coalescence picture of particle 
production in the intermediate $p_T$ range.

\section{Summary}
STAR has measured the  $K^{*0}$ resonance production at mid-rapidity in Au+Au 
and Cu+Cu collision systems at $\sqrt{s_{\mathrm{NN}}}$ = 62.4 GeV and 200 
GeV. A large sample of  Au+Au collision data at 
$\sqrt{s_{\mathrm{NN}}}$ = 200 GeV enables us to extend the measurements to 
$p_{\mathrm T}$ $\sim$ 5 GeV/$c$. The measured $dN/dy$ and 
$\langle p_{\mathrm T} \rangle$ of $K^{*0}$ are higher at $\sqrt{s_{\mathrm{NN}}}$ = 200 GeV 
compared to the corresponding values at $\sqrt{s_{\mathrm{NN}}}$ = 62.4 GeV.
For a given beam energy, the $dN/dy$ and 
$\langle p_{\mathrm T} \rangle$ are similar for Au+Au and Cu+Cu collisions at 
a given $\langle N_{\mathrm {part}} \rangle$. For $\sqrt{s_{\mathrm{NN}}}$ = 62.4 
and 200 GeV the $K^{*0}$ $\langle p_{\mathrm T} \rangle$ is comparable to 
the same for protons, indicating that the $\langle p_{\mathrm T} \rangle$ 
trends are dependent on the mass.

The $N(K^{*0})/N(K^{-})$ ratio in central Au+Au collisions at both 62.4 and 200 GeV 
is much smaller compared to the respective values in $p$+$p$ collisions. The data 
indicate that heavy ion collisions provide an environment with stronger re-scattering 
of $K^{*0}$ daughter particles relative to regeneration. The $N(\phi)/N(K^{*0})$ 
ratio in central Au+Au collisions at both 62.4 and 200 GeV is larger than 
that of $p$+$p$ collisions, again supporting the dominance of re-scattering
effects. The increase in the $N(\phi)/N(K^{*0})$ ratio as a function of beam energy 
and collision centrality also suggests strangeness enhancement in 
heavy-ion collisions. 

The large sample of Au+Au collision data at 
200 GeV allow for a quantitative estimation of elliptic flow of $K^{*0}$ 
and the interpretation of the $v_2$ in terms of a scaling based on the
number of constituent quarks. The results support the quark coalescence model
of particle production.
More explicitly, $K^{*0}$ are dominantly produced from direct quark combinations, 
with a negligible regenerated component. At low $p_{\mathrm T}$, the 
nuclear modification factor for $K^{*0}$ is seen to be similar for Au+Au 
collisions both at 62.4 and 200 GeV. At lower $p_{\mathrm T}$, $R_{CP}$  
for Au+Au collisions at $\sqrt{s_{\mathrm{NN}}}$ = 200 GeV is lower than
that for $\Lambda$ and $K^{0}_S$, which is consistent with the observation  
that the re-scattering effect dominates over regeneration effect. For $p_T >$ 
1.8 GeV/c, the $K^{*0}$ $R_{CP}$ in Au+Au collision at 200 GeV  more closely
follows that for $K^{0}_S$, and at the same time differs from that for $\Lambda$.
This also provides support for the quark coalescence picture at the intermediate
$p_{\mathrm T}$ ranges studied.

\section{Acknowledgments}
We thank the RHIC Operations Group and RCF at BNL, the NERSC Center at LBNL 
and the Open Science Grid consortium for providing resources and support. This 
work was supported in part by the Offices of NP and HEP within the U.S. DOE 
Office of Science, the U.S. NSF, the Sloan Foundation, the DFG cluster of 
excellence `Origin and Structure of the Universe' of Germany, CNRS/IN2P3, STFC and EPSRC 
of the United Kingdom, FAPESP CNPq of Brazil, Ministry of Ed. and Sci. of the 
Russian Federation, NNSFC, CAS, MoST, and MoE of China, GA and MSMT of the 
Czech Republic, FOM and NWO of the Netherlands, DAE, DST, and CSIR of India, 
Polish Ministry of Sci. and Higher Ed., Korea Research Foundation, Ministry of 
Sci., Ed. and Sports of the Rep. Of Croatia, Russian Ministry of Sci. and 
Tech, and RosAtom of Russia.

\end{document}